\documentclass[10pt,journal,  oneside, twocolumn]{IEEEtran}

\usepackage{multirow}
\usepackage{latexsym}
\usepackage{graphicx}
\usepackage{float}
\usepackage{amsmath}
\usepackage{amsthm}
\usepackage{lipsum}
\usepackage{subcaption}
\usepackage{graphicx}
\usepackage{authblk}
\usepackage{bm}
\usepackage{booktabs}
\usepackage{amsthm}
\usepackage[section]{placeins}
\usepackage{soul}
\usepackage{tabularx}

\usepackage{colortbl}

\usepackage{enumitem}

\usepackage{CJKutf8}

\usepackage[numbers,sort&compress]{natbib}

\makeatletter  
\newif\if@restonecol  
\makeatother

\usepackage[linesnumbered,ruled,vlined]{algorithm2e}
\usepackage{algpseudocode}  
\usepackage{amsmath}

\usepackage{amssymb}

\usepackage{mathrsfs}
\usepackage{subfig}
\usepackage{caption}
\newcounter{problem}

\renewcommand{\baselinestretch}{1.0}

\begin{document}

\title{Diverge-Merge Formation and MAC Control in Structured Airspace}


\author{Kai Xiong,~\IEEEmembership{Member,~IEEE}, Xingyu Wu, Ba Zhang, Li Wei,~\IEEEmembership{Member,~IEEE}, Min Zeng, Supeng Leng,~\IEEEmembership{Senior Member,~IEEE}


\thanks{

K. Xiong, X. Wu, B. Zhang, and S. Leng are with School of Information and Communication Engineering, University of Electronic Science and Technology of China, Chengdu, 611731, China.
}

\thanks{
L. Wei is with College of Information Science and Electronic Engineering, Zhejiang University, Hangzhou 310027, China.
}

\thanks{
M. Zeng is with the Network and Communication Research Institute, College of Information Engineering, Southwest University of Science and Technology, Mianyang, Sichuan 621010, China.
}



}


\maketitle

\begin{abstract}
The rapid scaling of advanced air mobility (AAM) makes corridor-based structured airspace a promising infrastructure for high-density unmanned aerial vehicle (UAV) traffic. 
Formation flight can improve corridor capacity by suppressing shockwave propagation, but rigid formations become inefficient or unsafe during ramp branching, merging, and congestion. 
To address this problem, this paper proposes a task-driven diverge-merge control framework for UAV formations in structured airspace. At the beginning, a corridor-ramp branching structured airspace model is established to characterize the traffic dynamics and spatial constraints.
Building upon this, a fast task-driven clustering mechanism integrates spatial connectivity, flight intent, and aerial task interactions to enable real-time diverge and merge for ramp branching and traffic reshaping.
To make the diverge-merge reconfigurations executable at the media access control (MAC) layer of the formation, a cluster-aware distributed time division multiple access (CAD-TDMA) protocol is further designed. It protects intra-cluster control synchronization while conservatively reusing low-risk inter-cluster slots.
Simulation results show that the proposed diverge-merge algorithm maintains near-zero geometrical misclassification under severe physical overlapping and congestion. With the formation diverge-merge traces, CAD-TDMA achieves the best delay--loss--throughput tradeoff over fixed TDMA and WiFi MAC.
It shows that the proposed formation control framework can jointly support real-time formation reconfiguration and reliable communication in corridor-ramp structured airspace.


\end{abstract}


\IEEEpeerreviewmaketitle

\section{Introduction}




\IEEEPARstart{A}{d}vanced air mobility (AAM) has catalyzed an unprecedented surge in low-altitude airspace utilization, positioning unmanned aerial vehicles (UAVs) as the primary carriers for next-generation urban logistics and on-demand transportation.
As aerial traffic scales toward density, the conventional paradigm of unconstrained free-flight becomes untenable, plagued by intractable route-crossing conflicts, congestion, and stringent regulatory safety mandates. 
To mitigate these systemic bottlenecks, corridor-based airspace has emerged as an indispensable architectural paradigm, organizing UAV flight into unidirectional, pipe-straight structures that enforce strict spatial boundaries and flow segregation, thereby laying the physical foundation for safe low-altitude operations.

\begin{figure}[htbp]
\centering
\includegraphics[width=0.98\columnwidth]{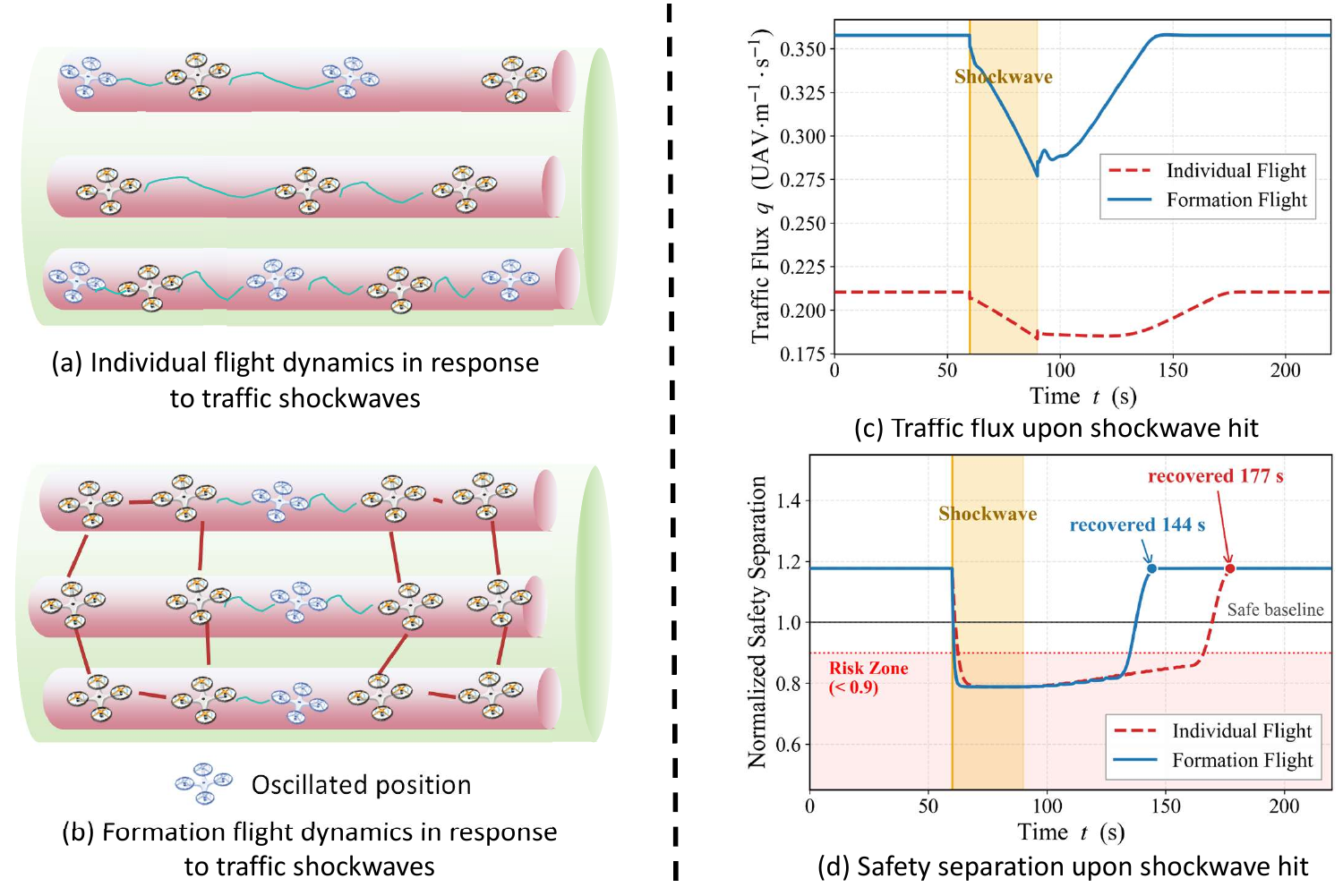} 
\caption{Corridor-based formation flight v.s. individual flight.}
\label{safety}
\end{figure}

Despite the structural advantages of corridors, independent UAV flight remains highly vulnerable to string shockwave effects, where minor velocity perturbations amplify backward and drastically degrade traffic flux. 
To breach this capacity limit, formation flight has been proposed as a highly effective paradigm. As illustrated in Fig.~\ref{safety}a, upon impact with a traffic shockwave, formation flight sustains and releases a higher traffic flux compared to individual flight. Furthermore, Fig.~\ref{safety}b shows that formation flight enables the swarm to recover the required safety separation faster after a shockwave disturbance. By restructuring multiple UAVs into a macroscopic kinematic unit, formation flight acts as a low-pass filter that suppresses high-frequency perturbations, thereby establishing the necessity of coordinated formations for high-capacity and safe corridor traffic.


\begin{figure}[!t]
\centering
\includegraphics[width=0.49\textwidth]{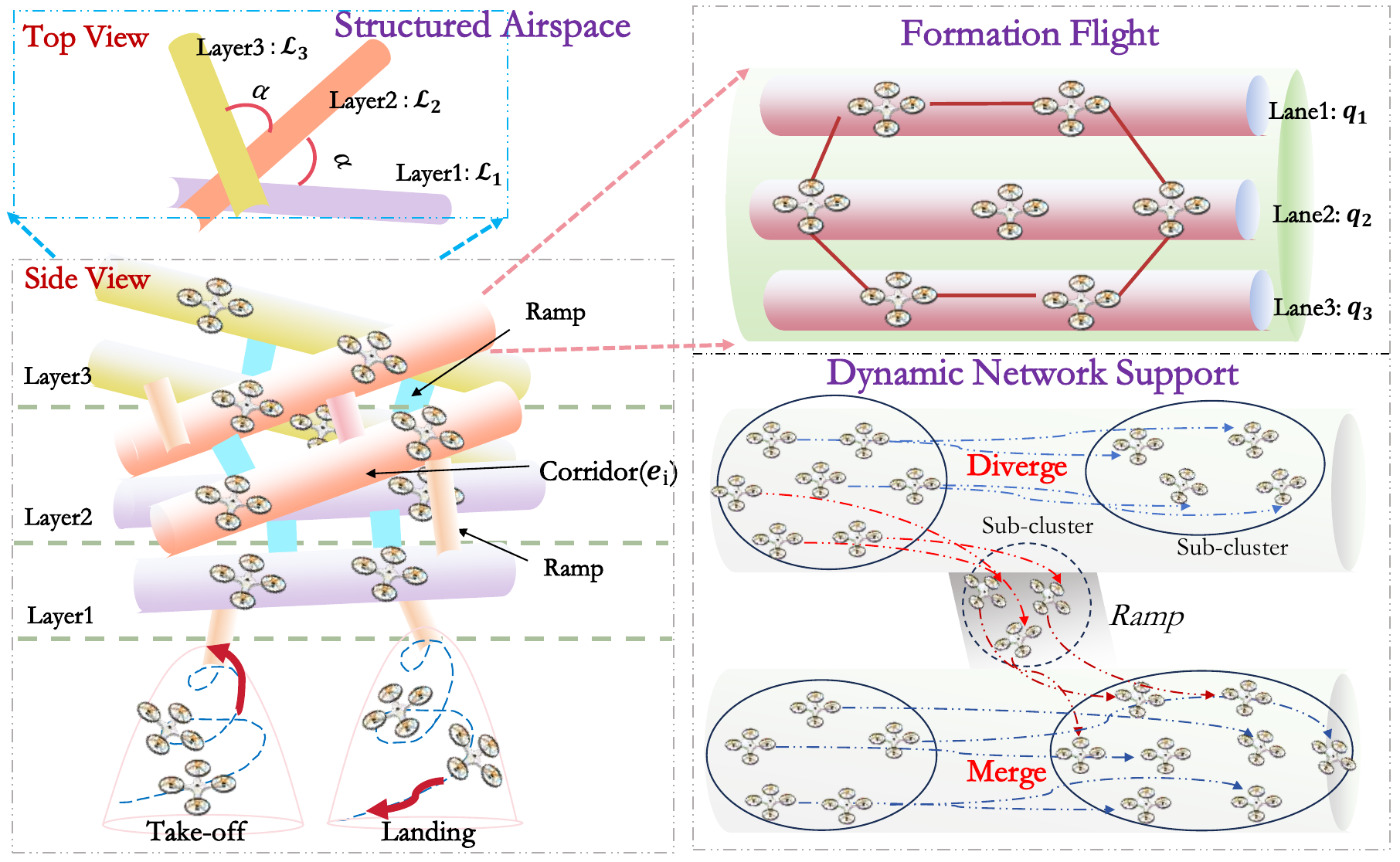} 
\caption{Corridor-ramp structured airspace.}
\label{fig:scenario_map}
\end{figure}

However, in realistic low-altitude traffic scenarios, UAVs cannot fly strictly straight. The reason is that navigating around no-fly zones, accommodating diverse destinations, and adapting to heterogeneous task distributions inevitably necessitate frequent changes in corridors with different directions. 
This demands a multidirectional corridor network. 
To guarantee flight safety, these multidirectional corridors cannot physically intersect at the same altitude. 
Consequently, to facilitate seamless direction switching among non-overlapping corridors, ramp structures, serving as dedicated transitional links between distinct corridors, are strictly required. 
Motivated by this, we propose the concept of a corridor-ramp structured airspace, as depicted in Fig.~\ref{fig:scenario_map}. This structured airspace comprises altitude-layered corridors interconnected by both intralayer and interlayer ramps, with individual corridors further longitudinally partitioned into multiple parallel lanes. It not only maintains corridor flight advantages but also provides a tractable environment for modeling complex route branching, merging, and capacity bottlenecks in high-density UAV dispatch.

Unfortunately, the introduction of the corridor-ramp structured airspace dictates that UAV formations must undergo dynamic diverge and merge maneuvers to navigate the network efficiently. 
As a formation approaches a ramp junction, its members often possess diverse downstream destinations, necessitating a diverge operation to decouple the macroscopic formation into sub-formations for corridor-switching navigation. Conversely, when sub-formations exit a ramp and enter a main corridor, maintaining overly fragmented formations wastes valuable spatial capacity and escalates routing overhead.
In such scenarios, these sub-formations must proactively execute a merge operation to re-aggregate with nearby co-directional formations, thereby maximizing the corridor's carrying density. Therefore, the capability to execute agile and appropriate diverge-merge transition is the prerequisite for efficient formation flight within this structured airspace.

Crucially, these transient diverge-merge operations impose severe stresses on the underlying UAV ad-hoc networks. Formation flight control, cooperative collision avoidance, and state synchronization heavily rely on reliable and stable intra-formation communications. 
During dynamic diverging and merging, formation node density fluctuates sharply, and formation-management traffic becomes highly bursty. Traditional medium access control (MAC) protocols struggle to provide comprehensive and stable network performance under such conditions. Contention-based protocols (e.g., CSMA/CA) suffer from severe air interface collisions and unpredictable delays when local density spikes, whereas fixed TDMA avoids random contention but introduces excessive waiting latency and lacks the flexibility to accommodate bursty, topology-aware control traffic. 
Consequently, a communication-support mechanism tailored for the dynamic diverge and merge process is needed to bridge the gap between physical formation control and cybernetic wireless access.

To address these intertwined challenges, this paper proposes a task-driven diverge-merge control framework tailored for the corridor-ramp structured airspace. First, we develop a task-driven diverge-merge algorithm that transcends traditional spatial-only clustering. This method integrates communication connectivity, 3-dimensional (3D) flight intent, and historical task interaction intensity. Furthermore, to support the resulting high-frequency network changes at the wireless access layer, we design a cluster-aware distributed TDMA (CAD-TDMA) protocol. CAD-TDMA explicitly maps the task-aware cluster labels into distributed frame-level slot management. It strictly protects intra-cluster owner slots for critical control synchronization while conservatively reusing low-risk inter-cluster slots based on passive overhearing and local risk history, thereby achieving an optimal delay-loss-throughput tradeoff.
Specifically, the main contributions of this paper are summarized as follows:


\begin{itemize}

\item We propose a corridor-ramp structured airspace model that explicitly captures the geometric constraints and topological complexities of low-altitude traffic. By incorporating altitude-layered, multi-lane unidirectional corridors interconnected by ramps, this structured airspace enforces flow segregation to eliminate free-flight conflicts while providing a foundation for modeling capacity bottlenecks during high-density UAV dispatch.


\item We develop a task-driven dynamic diverge-and-merge collaborative control algorithm that achieves precise formation reconfiguration under stringent spatial constraints. Distinct from traditional group partitioning, our approach leverages unsupervised state-feedback for closed-loop weight adaptation and an accelerated Fast Spectral Clustering (FSC) technique. By anchoring topology evolution on historical task interactions and 3D flight intent, the algorithm effectively prevents blind fragmentation and erroneous merging when physical distance cues degrade during complex ramp maneuvers.

\item We design a cluster-aware distributed TDMA (CAD-TDMA) MAC protocol specifically tailored to support the bursty communication demands of formation diverge-and-merge operations. By mapping task-aware cluster labels into frame-level slot management, CAD-TDMA protects intra-cluster reserved slots for critical synchronization and conservatively reuses low-risk inter-cluster slots according to local observation history. Extensive ns-3 validations demonstrate that this communication-support mechanism achieves a more balanced delay--reliability tradeoff than fixed TDMA and WiFi baselines in dense, dynamic corridor flight scenarios.

\end{itemize}

The remainder is organized as follows: Sec. II reviews related work. Sec. III presents the system model. Sec. IV and V give the proposed algorithms. Sec. VI presents the simulation, and Sec. VII concludes the paper.

\section{Related Work}
The realization of AAM necessitates the seamless integration of airspace architecture, flight formation coordination, and wireless communication support. 
This section reviews the state-of-the-art in these three interconnected domains and identifies the critical gaps that motivate our diverge-merge formation control and network design.

\subsection{Structured airspace architecture}

As the scale of AAM expands, the conventional free-flight paradigm becomes nonviable for managing high-density UAVs due to massive route-crossing conflicts and collision risks~\cite{10718279}.
Consequently, both academia and aviation authorities have proposed various airspace architectures to regulate low-altitude operations~\cite{KIM20261}. 
Existing low-altitude airspace studies have explored static grid-based partitioning via geofencing~\cite{cheaitouOptimizingVisualizingDynamic2026} and multi-layered airspace designs aimed at vertical flow segregation~\cite{11297260}. 
Additionally, directional corridors championed by the federal aviation administration (FAA)~\cite{faa_uam_conops_2023,9447255} and cooperative virtual-tube models~\cite{9591313} have been extensively investigated. 
These pioneering airspace architectures have advanced aerial traffic management, capacity estimation, and route planning.

Despite exploring diverse airspace architectures, existing studies overlook the extreme spatial squeezing effects imposed by geometric boundaries on formation flight at the structured bottlenecks. 
They fail to address the network-physical coupling between macroscopic spatial constraints and microscopic formation control and communication scheduling.


\subsection{Formation flight control}
In structured airspace, formation flight is a promising paradigm for enhancing macroscopic traffic flux and mitigating collision risks associated with uncoordinated movements. However, maintaining rigid formations is impractical when confronted with narrow ramps and dynamic task requirements. Hence, dynamic formation reshaping has emerged as a prominent research area~\cite{10839946, 11011677, 10174704}.
Existing literature explores formation diverge-merge mechanisms: some employ autonomous decision, such as improved potential field control or multi-agent consensus protocols, to achieve adaptive obstacle avoidance~\cite{11086501,11048703}. Moreover, graph-theoretic clustering and heuristic rules are widely utilized. By partitioning adjacency matrices based on spatial metrics and communication quality to sever weak edges, these methods facilitate safe formation diverging at route convergences~\cite{10470428,9486931,10960266}.

Despite these efforts, current strategies exhibit limitations. Traditional algorithms often rely on homogeneous physical distance or relative velocity features, neglecting historical task interactions and flight intents. Furthermore, handling real-time evolution for large-scale nodes incurs prohibitive computational overhead for eigen-decomposition in spectral clustering, failing to meet millisecond-level low-altitude operation requirements~\cite{11429577,8281623}.

\subsection{Distributed MAC protocols}

MAC design has been widely studied in wireless Ad-Hoc, vehicular, and aerial networks. 
Classical carrier-sense random access protocols, such as carrier sense multiple access with collision avoidance (CSMA/CA) and IEEE 802.11 distributed coordination function (DCF), provide flexible distributed access without global coordination. 
However, their performance is strongly affected by contention intensity and hidden-terminal conditions. 
In dense and dynamic UAV formations, diverge-merge operations may cause sudden node-density variations and bursty traffic, increasing collision probability, retransmissions, and access delay uncertainty~\cite{gupta2016survey,mozaffari2019tutorial,li2016energy}.
In contrast, TDMA-based MAC divides time into frames and slots, which can provide more predictable transmission opportunities by reducing random contention.
It is suitable as a MAC-support mechanism for high-dynamic formations.

Existing distributed TDMA MAC protocols have been investigated in Ad-Hoc, vehicular, and aerial networks. 
For example, distributed slot reservation and reuse have been widely studied for reliable broadcast and delay-sensitive communication~\cite{omar2013vemac,hadded2015tdma,zhang2015centralized,jiang2016ptmac,satmac2022}.
Nevertheless, these protocols mainly manage slots according to node-level reservations, local topology, or traffic demand. 
They do not explicitly apply the task-aware cluster labels, and therefore cannot directly distinguish intra-/inter-cluster synchronization slots.



\section{System Model}
\label{sec:system_model}

Conventional swarm control isolates formation reshaping from network scheduling, inevitably triggering disconnections and collisions.
To fulfill the research gap, this section provides a joint formation and wireless MAC scheduling optimization model. 
Specifically, we first construct the corridor-ramp airspace and UAV kinematics. 
Then, a multi-dimensional similarity metric is proposed to optimize the formation control.

\subsection{Corridor-ramp structured airspace}
Emerging urban traffic management frameworks such as UTM
\cite{kopardekar2016unmanned} and U-space \cite{sesar2017u} confine UAV operations within corridors. We accordingly propose a multi-layered and multi-lane corridor-ramp structured airspace.
The airspace is stratified into $N_{L}$ altitude layers
$\{\mathcal{L}_1,\dots,\mathcal{L}_{N_{L}}\}$ that separate conflicting flows.
Each layer $\mathcal{L}_{\ell}$ hosts a set of parallel unidirectional corridors $\{e_1, e_2, \dots\}$, and adjacent layers differ by a fixed heading angle $\alpha$ so that their flows do not intersect at the same altitude, as shown in the top view of Fig.~\ref{fig:scenario_map}.
Each corridor $e$ is in turn divided into $n_{L,e}$ parallel lanes $\{q_1,\dots,q_{n_{L,e}}\}$ across its cross-section, which form the discrete lateral slots along which a formation arranges its members, as shown in the formation flight panel of Fig.~\ref{fig:scenario_map}.
Corridors are interconnected by ramps. A ramp is a transitional link that connects either two corridors within the same layer or two corridors residing in different layers, so that a UAV switches corridors through a ramp according to its destination and the downstream congestion.
The airspace is abstracted as a directed graph $\mathcal{G}=(\mathcal{V},\mathcal{E})$, where $\mathcal{V}$ collects the junctions such as waypoints and ramp endpoints, and each edge $e\in\mathcal{E}$ is a corridor segment:
\begin{equation}
    e = \big( l_e,\, R_e,\, n_{L,e},\, z_e,\, C_{\max,e}(v) \big),
    \label{sdakjbveh}
\end{equation}
\noindent with length $l_e$, cross-sectional radius $R_e$, lane count $n_{L,e}\le\lfloor 2R_e/w_L\rfloor$, lateral lane spacing $w_L$, and host layer $z_e\in\{1,\dots,N_{L}\}$ determining corridor altitude and heading.
Rear-end safety within each lane $q$ enforces a velocity-dependent longitudinal separation $d_{safe}(v)=d_0+\tau v+\frac{v^2}{2a_{\max}}$, which combines the static margin $d_0$, the reaction delay $\tau$, and the maximum deceleration $a_{\max}$. Aggregating the $n_{L,e}$ lanes, the corridor capacity is:
\begin{equation}
    C_{\max,e}(v) = n_{L,e}\left\lfloor \frac{l_e}{d_{safe}(v)} \right\rfloor .
\end{equation}
\noindent When the speed rises or a formation approaches a ramp, both the per-lane admissible count and the available lane number decrease, so a formation executes diverge-merge maneuvers to change the lane occupancy across different corridors and keep the traffic load within the corridor capacity $C_{\max,e}(v)$.

\subsection{Formation flight dynamics}
Conventional free-flight mobility models fail to capture the severe physical mutual exclusions inherent in confined structured airspace. 
To reflect fluid-like congestion and collision avoidance, the kinetic state of UAV $u_i \in \mathcal{U}$ at time $t$ is assembled by its position $\mathbf{p}_i(t) \in \mathbb{R}^3$ and velocity $\mathbf{v}_i(t) \in \mathbb{R}^3$.

Let $\mathbf{p}_{tgt,i}$ denote the terminal waypoint for node $u_i$. 
To enforce the required safety separation $d_{safe}$ during spatial squeezing, we integrate an artificial potential field (APF)-based spatial repulsive force. For the detectable neighbor set $\mathcal{N}_i$, the repulsion $\mathbf{f}_{rep,i}(t)$ is activated exclusively when the inter-UAV distance $d_{i,j}(t) = \|\mathbf{p}_i(t) - \mathbf{p}_j(t)\|$ breaches $d_{safe}$:
\begin{equation}
\mathbf{f}_{rep,i}(t) = \!\!\!\!\!\!\!\!\sum_{j \in \mathcal{N}_i, d_{i,j} < d_{safe}} \!\!\!\!\!\!\!\!\kappa \left( \frac{1}{d_{i,j}(t)} - \frac{1}{d_{safe}} \right) \frac{\mathbf{p}_i(t) - \mathbf{p}_j(t)}{d_{i,j}(t)},
\label{eq:repulsion}
\end{equation}
\noindent where $\kappa$ is the repulsive gain. 
The actual steering $\mathbf{u}_i(t)$ synthesizes destination attraction, velocity damping, and this collision repulsion. To comply with the physical limits, the joint force is bounded by the allowable acceleration $a_{max}$:
\begin{equation}
\mathbf{u}_i(t) \!=\! \operatorname{sat}_{a_{max}} \!\! \left( \! v_{max} \frac{\mathbf{p}_{tgt,i} \!-\! \mathbf{p}_i(t)}{\|\mathbf{p}_{tgt,i} \!-\! \mathbf{p}_i(t)\|} \!-\! \mathbf{v}_i(t) \!+\! \mathbf{f}_{rep,i}(t) \!\right),
\label{eq:control_law}
\end{equation}

\noindent where the standard vector saturation operator is $\operatorname{sat}_{C}(\mathbf{x}) = \mathbf{x} \min(1, \frac{C}{\|\mathbf{x}\|})$. Besides, navigating narrow corridors exacerbates aerodynamic disturbances (e.g., wall effects). This exogenous perturbation is modeled as a Gaussian noise $\mathbf{n}_i(t) \sim \mathcal{N}(0, \sigma^2 \mathbf{I})$. The discrete-time kinematic updates, bounded by the allowable speed $v_{max}$, are given as:
\begin{equation}
\begin{split}
\mathbf{v}_i(t+1) &= \operatorname{sat}_{v_{max}}\Big(\mathbf{v}_i(t) + \mathbf{u}_i(t)\Delta t\Big) + \mathbf{n}_i(t),\\
\mathbf{p}_i(t+1) &= \mathbf{p}_i(t) + \mathbf{v}_i(t+1)\Delta t.
\label{eq:position_update}
\end{split}
\end{equation}
Ultimately, this mobility model couples kinematic bottlenecks with fluidic congestion, providing a physical foundation for evaluating the subsequent formation transition mechanisms.

\subsection{Multi-dimensional clustered similarity}

Conventional spatial clustering algorithms blindly equate physical proximity with logical affiliation. This is a fatal assumption in structured airspace where functionally distinct formations frequently cross corridors and ramps. 
To shatter this spatial rigidity, we construct a time-varying, undirected, and non-negative logical graph $\mathcal{H}(t) = (\mathcal{U}, \mathbf{S}(t))$ to represent the cyber-physical relationships within the formation. The similarity matrix $\mathbf{S}(t) \in \mathbb{R}^{N \times N}$ quantifies the logical coupling degree between any pair of nodes. 
To satisfy the symmetric and non-negative prerequisites for the subsequent Normalized Cut (Ncut) operations, $\mathbf{S}(t)$ is assembled by three density-adaptive similarities, which are the communication similarity, flight intent similarity, and task similarity.

\subsubsection{Communication similarity}
This similarity characterizes the communication link reliability. In congested spaces (e.g., rear-end congestion in a single corridor or ramp merging), the abrupt surge in UAV density leads to severe communication interference. Assuming the transmission power of all UAVs is $P_{tx}$. When node $j$ receives a signal from node $i$, the physical-layer Signal-to-Interference-plus-Noise Ratio (SINR) is~\cite{8918497}:
\begin{equation}
\gamma_{i,j}(t) = \frac{P_{tx} G_0 d_{i,j}^{-\alpha}(t) |h_{i,j}|^2}{N_0 + \sum_{k \in \mathcal{U} \setminus \{i,j\}} P_{tx} G_0 d_{k,j}^{-\alpha}(t) |h_{k,j}|^2},
\end{equation}
\noindent where $d_{i,j}(t)$ is the distance between UAV $i$ and $j$, $\alpha$ is the path loss exponent, $G_0$ is the reference channel gain, $h$ is the fading coefficient, and $N_0$ is the ambient noise power. 
Due to the distinct interference perceived at different UAVs (receivers), we then define the communication similarity using a Sigmoid activation function that maps the averaged SINR of bidirectional UAV $i$ and $j$ into a continuous probability:
\begin{equation}
    S_{\mathcal{L}}^{i,j} = \frac{1}{1 + \exp\left(-\kappa \left(\frac{\gamma_{i,j}(t) + \gamma_{j,i}(t)}{2} - \gamma_{th}\right)\right)},
\end{equation}
\noindent where $\gamma_{th}$ is the minimum SINR demodulation threshold, and $\kappa$ controls the activation steepness. This formulation intrinsically filters out weak links while preserving symmetric topological connectivity.

\subsubsection{Flight intent similarity}
During UAV formation merging and overtaking, distinct task groups may spatially overlap. Flight intent serves to decouple spatially intermingled UAVs. The intent similarity $S_{\mathcal{I}}^{i,j} \in [0,1]$ is jointly determined by short-term kinematic heading alignment and long-term 3D trajectory convergence:
\begin{equation}
\begin{split}
    S_{\mathcal{I}}^{i,j} &= \lambda \left( \frac{1}{2} + \frac{\mathbf{v}_i^T(t) \mathbf{v}_j(t)}{2 \|\mathbf{v}_i(t)\| \|\mathbf{v}_j(t)\|} \right) \\
    &\quad + (1-\lambda) \exp\left(-\frac{\|\mathbf{p}_{tgt,i} - \mathbf{p}_{tgt,j}\|^2}{2\sigma_{tgt}^2}\right),
\end{split}
\end{equation}
where $\mathbf{v}(t)$ denotes the instantaneous velocity vector, $\mathbf{p}_{tgt}$ represents the 3D terminal task waypoint, $\lambda \in [0, 1]$ balances the influence of short-term heading and long-term destination, and $\sigma_{tgt}$ defines the distance threshold for destination similarity. 
The first term calculates the cosine similarity of velocity vectors to align UAV headings, while the second term employs a Gaussian kernel to measure terminal waypoint proximity.

\subsubsection{Task interaction similarity}
To verify the underlying logical affinity, we mine the historical task interactions through the proposed time-decayed frequent pattern tree (TD-FPTree) algorithm. Denote $\Omega_{i,j}(\tau) \in \{0,1\}$ as a binary indicator, where $\Omega_{i,j}(\tau) = 1$ if node $i$ and node $j$ cooperate in the same task transaction at time $\tau$, and $0$ otherwise. To enhance recent interactions while fading obsolete relations, the interaction intensity $S_{\mathcal{T}}^{i,j}$ is defined over a sliding time window $[t-T_w, t]$:
\begin{equation}
    S_{\mathcal{T}}^{i,j} = \frac{1}{\Phi} \sum_{\tau=t-T_w}^{t} \Omega_{i,j}(\tau) e^{-\eta_d(t-\tau)},
\end{equation}

\noindent where $\eta_d > 0$ is the coefficient, and $T_w$ is the window length. Crucially, the normalization factor $\Phi = \sum_{\tau=t-T_w}^{t} e^{-\eta(t-\tau)}$ bounds $S_{\mathcal{T}}^{i,j}$ within $[0,1]$.
Finally, the adjacency matrix of the logical connection graph $\mathcal{H}(t)$ is a weighted sum of three similarity metrics:
\begin{equation}
S^{i,j}(t) = \beta_1(t) S_{\mathcal{L}}^{i,j} + \beta_2(t) S_{\mathcal{I}}^{i,j} + \beta_3(t) S_{\mathcal{T}}^{i,j},
\end{equation}

\noindent where $\beta_k(t) \ge 0$ and $\sum_i\beta_i(t) = 1$. This construction ensures that $\mathbf{S}(t)$ is symmetric, non-negative, and properly bounded within $[0,1]$, satisfying all necessary and sufficient prerequisites for the subsequent graph partitioning operations.

\subsection{Problem formulation}
Upon the multi-dimensional connection graph $\mathcal{H}(t) = (\mathcal{U}, \mathbf{S}(t))$, the dynamic diverge-and-merge process is formed as a capacity-constrained Ncut problem. Let the formation $\mathcal{U}$ be partitioned into $k$ disjoint clusters (sub-formations), denoted as $\mathcal{C} = \{C_1, C_2, \dots, C_k\}$.
Furthermore, the objective is designed to maximize the intra-cluster network-formation cohesiveness while safely severing weak logical connections. 
For any given cluster $C_m$, the external edge-cut $ \sum_{i \in C_m} \sum_{j \notin C_m} S^{i,j}(t)$ (overall severed logical connections) and the degree-volume $\sum_{i \in C_m} \sum_{j \in \mathcal{U}} S^{i,j}(t)$ (overall connectivity intensity) are functions of similarity matrix $\mathbf{S}(t)$.

Unlike traditional clustering algorithms, any formation control in structured airspace must respect the underlying spatial and communication bottlenecks. 
To prevent oversized formations from overwhelming the MAC-layer scheduling, the partition is subjected to a wireless access capacity $N_{\max}$, dictated by the available air-interface resources. 
This constraint acts as the bridge linking the formation control with wireless limits. Consequently, the optimization problem is:
\begin{equation}
\begin{gathered}
\textbf{P1}: \ \  \min_{\mathcal{C}} \!\!\!\! \quad
\sum_{m=1}^{k}
\frac{\sum_{i \in C_m} \sum_{j \notin C_m} S^{i,j}(t)}
     {\sum_{i \in C_m} \sum_{j \in \mathcal{U}} S^{i,j}(t)} \\
\mathrm{s.t.}\quad
|C_m|\le N_{\max},\quad \forall m \in \{1,\ldots,k\}.
\end{gathered}
\label{eq:ncut_problem}
\end{equation}



\section{Dynamic Diverge-Merge Control}
Solving the network-spatial constrained cluster partitioning problem \textbf{P1} in real time constitutes the core challenge of the proposed diverge-merge control.
To meet the stringent real-time computational demands imposed by high-frequency formation transitions, the proposed algorithm is designed into three core modules: 1) an on-demand diverge-and-merge mechanism; 2) a fast spectral clustering paradigm; and 3) a closed-loop adaptive clustering adjustment. 

\subsection{Network-supported formation control}

\begin{figure}[htbp]
\centering
\includegraphics[width=0.98\columnwidth]{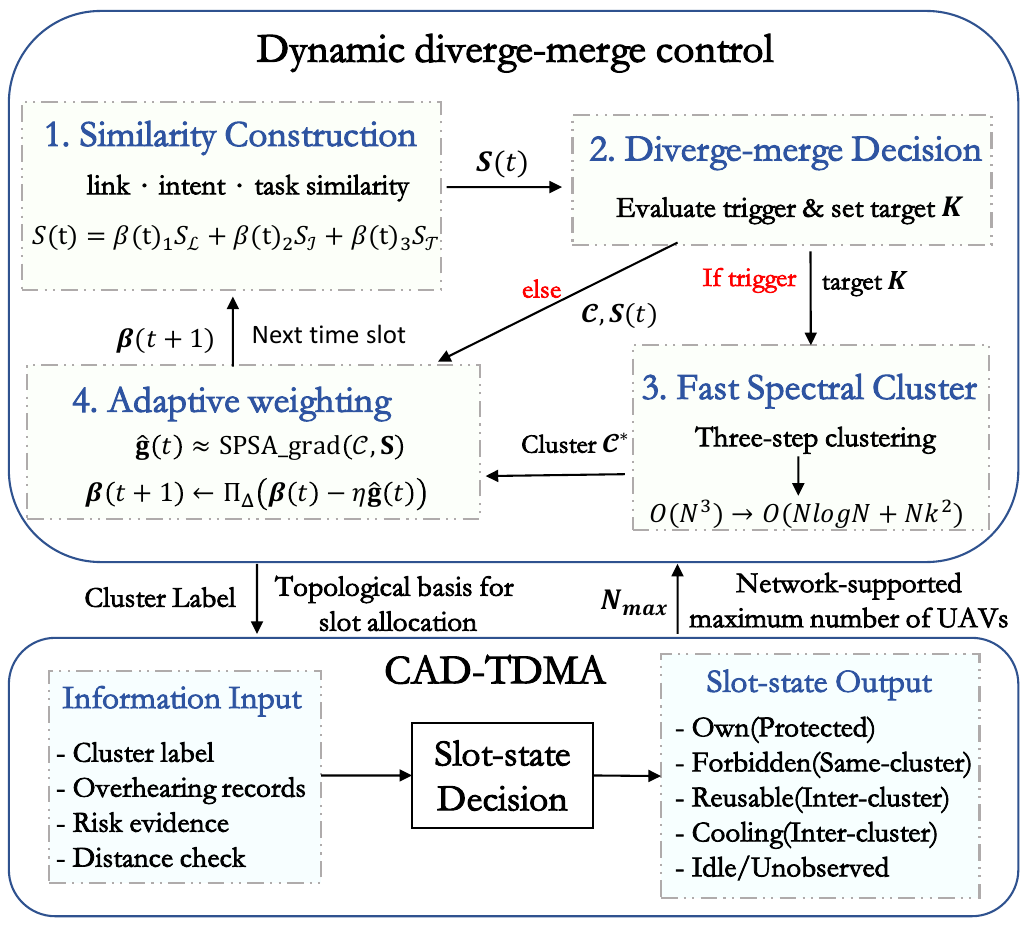} 
\caption{Network-supported diverge-merge control framework.}
\label{Flow}
\end{figure}

To shatter the network-formation isolation inherent in swarm optimization, we propose a network-support diverge and merge formation control scheme, as depicted in Fig.~\ref{Flow}.
Specifically, this architecture has two coupled modules. At the upper module, the dynamic diverge-merge control operates through a four-step closed loop. Rather than relying on spatial proximity, the process begins by evaluating multi-dimensional connections (communication link, flight intent, and task) to construct the similarity matrix. A subsequent formation control component evaluates trigger conditions to set the target cluster count. If the triggering conditions are met, the formation performs a FSC algorithm to reshape its envelope geometry; otherwise, it bypasses re-clustering. Finally, an adaptive weighting component computes the weighting parameters for the next time slot, providing feedback to the similarity construction. 

At the lower module of Fig.~\ref{Flow}, the generated cluster labels from the diverge-merge control module serve as the basis for MAC slot allocation. CAD-TDMA maps these labels, alongside passive overhearing records, risk evidence, and distance checks, into local slot-state decisions. 
Each slot is classified as one of the Own (Protected), Forbidden (Same-cluster), Reusable (Inter-cluster), Cooling (Inter-cluster), or Idle/Unobserved, ensuring same-cluster slots are protected while low-risk inter-cluster slots can be reused. 
To maintain network-formation consistency, CAD-TDMA feeds back the wireless access capacity to the diverge-merge control module of Fig.~\ref{Flow}, ensuring that the number of cluster and an individual cluster scale adhere to the real-time wireless access capability.

\subsection{Dynamic diverge-and-merge triggers}

In spatial confined corridors, relying on a static number of clusters $k$ inevitably degrades communication and collision-proof flight efficacy. To avoid this, we design an event-driven mechanism to dynamically manage the formation scale.

As multiple clusters converge at spatial bottlenecks (e.g., multi-ramp junctions), inter-cluster distances drastically diminish. To accurately capture imminent collision risks, the physical inter-cluster distance is defined by the minimum edge distance: $D_{inter}(C_m, C_n) = \min_{i \in C_m, j \in C_n} d_{i,j}(t)$. 
A merge maneuver is executed when $D_{inter}$ falls below a predefined threshold $d_{merge}$. 
Crucially, in structured airspace, physical survival strictly supersedes logical task independence. 
Crucially, in structured airspace, flight safety supersedes logical task independence. 
Therefore, physically proximal clusters are compelled to logically merge, forcing them to share a unified TDMA frame for coordinated collision avoidance. Provided their merged cluster size satisfies the MAC access capacity, the merge trigger is formulated as:
\begin{equation}
\label{eq:merge_trigger}
D_{inter}(C_m, C_n) \leq d_{merge} \quad \wedge \quad |C_m| + |C_n| \leq N_{\max}.
\end{equation}
\noindent Once the triggering conditions are met, the adjacent formations logically merge ($k \leftarrow k - 1$).

Conversely, maintaining an oversized formation incurs prohibitive intra-communication delays. To comply with the MAC access capacity $N_{\max}$ limitation, the local degradation loss is evaluated through the maximum physical dispersion:
$\epsilon_{\mathcal{L}}(C_m) = \max_{i,j \in C_m} ( d_{i,j}(t) / R_c )^2$. Hence, the local diverge is activated if:
\begin{equation}
\label{eq:local_diverge}
|C_m| > N_{\max} \quad \vee \quad \epsilon_{\mathcal{L}}(C_m) \geq \epsilon_{th}.
\end{equation}

Upon activation, the FSC algorithm is locally invoked to cleave this specific oversized formation ($k_{local} \leftarrow \lceil |C_m| / N_{\max} \rceil$), releasing network pressure.

Prolonged reliance on local operations may gradually fragment the macro-connections. Let $k_{\max} \geq \lceil |\mathcal{U}| / N_{\max} \rceil$ denote the upper bound of manageable sub-nets for the corridor routing protocol. When the total cluster count exceeds this threshold ($k > k_{\max}$), local adjustments are deemed insufficient. Serving exclusively as a last-resort fallback due to its high computational overhead, a cluster-wide FSC execution is triggered ($k_{global} \leftarrow \lceil |\mathcal{U}| / N_{\max} \rceil$) to eradicate accumulated formations fragmentation and restore traffic efficiency.

\subsection{Fast spectral clustering}

While spectral clustering provides mathematically optimal topology boundaries, its baseline $O(N^3)$ complexity is computationally catastrophic for resource-constrained UAVs executing high-frequency maneuvers. To shatter this bottleneck, we propose a three-stage FSC pipeline that drastically curtails the computational footprint without sacrificing algebraic rigor.

First, we aggressively sparsify the dense similarity matrix $\mathbf{S}(t)$ to bypass the $O(N^2)$ graph construction bottleneck. By deploying spatial KD-trees, we construct a mutual $K$-nearest neighbor ($K$-NN) graph in just $O(N \log N)$ time. For any node $i$ with its $K_{nn}$ strongest connections denoted by $\mathcal{N}_K(i)$, the sparse symmetric adjacency matrix is defined as:
\begin{equation}
S_{sparse}^{i,j} = 
\begin{cases}
S^{i,j}(t), & \text{if } j \in \mathcal{N}_K(i) \text{ or } i \in \mathcal{N}_K(j) \\
0, & \text{otherwise}
\end{cases}.
\end{equation}

\noindent To prevent ill-conditioned matrix exceptions caused by isolated nodes during sparsification, a tiny numerical stabilizer $\delta > 0$ is injected. The regularized symmetric normalized Laplacian is given as:
\begin{equation}
\mathbf{L}_{sym}(t) = \mathbf{I} - \mathbf{D}_{\delta}^{-1/2} \mathbf{S}_{sparse}(t) \mathbf{D}_{\delta}^{-1/2},
\end{equation}
where $\mathbf{D}_{\delta} = \mathbf{D}_{sparse}(t) + \delta \mathbf{I}$. This strict regularization securely bounds the eigenvalues within $[0, 2]$.

\begin{algorithm}
\caption{Fast Spectral Clustering (FSC)} \label{alg:FSC}
\textbf{Input}: $\mathbf{S}(t)$, neighbor count $K_{nn}$, numerical shift $\delta$\\ 
\textbf{Output}: Cluster partition $\mathcal{C} = \{C_1, C_2, \dots, C_k\}$

Sparsify $\mathbf{S}(t)$ by mutual $K_{nn}$ nearest neighbors \\
Regularize degree matrix: $\mathbf{D}_{\delta} \gets \mathrm{diag}(\mathbf{S}_{sparse}\mathbf{1}) + \delta \mathbf{I}$ \\
Compute symmetric normalized Laplacian: $\mathbf{L}_{sym} \gets \mathbf{I} - \mathbf{D}_{\delta}^{-1/2} \mathbf{S}_{sparse} \mathbf{D}_{\delta}^{-1/2}$

Extract top $k$ eigenvectors of $\mathbf{L}_{sym}$ via IRLM \\
Normalize eigenvector matrix $\mathbf{U}$ by rows to get embedding $\mathbf{E}^* \in \mathbb{R}^{N \times k}$

Apply Mini-Batch K-Means++ on $\mathbf{E}^*$  \\
\end{algorithm}

Second, rather than executing an exhaustive $O(N^3)$ eigendecomposition, we employ the Implicitly Restarted Lanczos Method (IRLM) \cite{10.1137/0613025}. By extracting only the first $k$ eigenvectors corresponding to the smallest eigenvalues, IRLM drastically curtails the embedding complexity to $O(N k^2)$. The optimal cluster count $k$ is dynamically determined by the standard eigengap heuristic, $k = \arg\max_i (\lambda_{i+1} - \lambda_i)$. The extracted eigenvectors are then row-wise $\ell_2$-normalized to form the low-dimensional feature matrix $\mathbf{E}^* \in \mathbb{R}^{N \times k}$.

Finally, to accelerate the discretization phase, we replace standard K-Means with MiniBatchKMeans on $\mathbf{E}^*$. Initialized via KMeans++~\cite{2007K}, it updates centroids using random subsets rather than the entire swarm, operating in just $O(b \cdot k \cdot I)$ where $b$ is the batch size and $I$ is the iteration count. 

By cascading KD-tree sparsification ($O(N \log N)$), IRLM partial extraction ($O(N k^2)$), and MiniBatch discretization ($O(b \cdot k \cdot I)$), the computational core of the FSC mechanism is strictly bounded by $O(N \log N + N k^2)$. Since the batch size, iterations, and cluster count satisfy $b, I, k \ll N$, this FSC mechanism solidly guarantees compliance with the strict real-time processing constraints of onboard flight controllers, with the complete procedure given in Alg.~\ref{alg:FSC}.


\subsection{Adaptive weight adjustment}
Static topology weights leave the swarm blind to structural degradation in fluid corridor environments. The three similarity modalities also differ in how informative they are over time. In free flow, spatial proximity separates formations cleanly, while inside a congested corridor the formations interpenetrate and the link modality loses its discriminative power. The weighting vector $\boldsymbol{\beta}(t) = [\beta_1(t), \beta_2(t), \beta_3(t)]^T$ is therefore adapted online via structural feedback. We define three unsupervised penalty metrics to quantify distinct topological defects:

\subsubsection{Communication degradation}
This metric evaluates the loss of intra-cluster spatial cohesiveness by measuring physical link dispersion. Let $R_c$ denote the maximum reliable communication range. The global penalty is formulated as the mean squared distance ratio:
\begin{equation}
\epsilon_{\mathcal{L}}(t) = \frac{1}{k} \sum_{m=1}^k \frac{1}{|C_m|^2} \sum_{i \in C_m} \sum_{j \in C_m} \left( \frac{d_{i,j}(t)}{R_c} \right)^2.
\end{equation}
An elevated $\epsilon_{\mathcal{L}}(t)$ indicates that the partition stretches intra-cluster links beyond reliable range.

\subsubsection{Clustering oscillation}
To penalize structural jitter while bypassing the label permutation problem of unsupervised clustering, we introduce a co-membership indicator matrix $\mathbf{Z}(t) \in \{0,1\}^{N \times N}$, where $Z_{i,j}(t) = 1$ if UAVs $i$ and $j$ share a cluster and $0$ otherwise. The macroscopic oscillation is quantified by the temporal disagreement:
\begin{equation}
\epsilon_{\mathcal{I}}(t) = \frac{1}{N(N-1)} \sum_{i \in \mathcal{U}} \sum_{j \neq i} \left| Z_{i,j}(t) - Z_{i,j}(t-1) \right|.
\end{equation}
A high $\epsilon_{\mathcal{I}}(t)$ indicates that the partition churns between slots instead of tracking a persistent structure.

\begin{algorithm}
\caption{Fast Diverge-and-Merge Control} \label{Alg:DynamicControl}

\textbf{Input}: $\mathcal{U}, N_{\max}, k_{\max}, \epsilon_{th}, d_{merge}, d_{diverge}, \eta, T_{\beta}$ \\
\textbf{Output}: Cluster assignments $\mathcal{C}(t)$

Initialize $\boldsymbol{\beta}(0) \gets [1/3, 1/3, 1/3]^T$

\For{each time slot $t$} {
    \If{$t \bmod T_{\beta} = 0$} {
        Update $\boldsymbol{\beta}(t)$ using SPSA gradient $\hat{\mathbf{g}}(t)$
    }
    
    Refresh $\mathbf{S}(t)$ using $\boldsymbol{\beta}(t)$, set $k_{cap} \gets \lceil |\mathcal{U}| / N_{\max} \rceil$

    \If{Diverge Condition } {
        $k_{tgt} \gets k(t-1) + 1$
    } \ElseIf{Merge Condition } {
        $k_{tgt} \gets k(t-1) - 1$
    } \Else {
        $k_{tgt} \gets \operatorname{EigenGap}(\mathbf{S}(t))$
    }

    $k_{tgt} \gets \operatorname{clip}(k_{tgt}, \max(k_{\min}, k_{cap}), k_{\max})$

    \If{$k_{tgt} \neq k(t-1)$ OR Trigger conditions met} {
        $\mathcal{C}(t) \gets \operatorname{FSC}(\mathbf{S}(t), k_{tgt})$ \hfill \tcp*{via Alg.~\ref{alg:FSC}}
    } \Else {
        $\mathcal{C}(t) \gets \mathcal{C}(t-1)$
    }
}
\end{algorithm}

\subsubsection{Logical fragmentation}
This metric quantifies the proportion of collaborative relationships severed by the current partition, defined as the normalized cut of the historical interaction graph:
\begin{equation}
\epsilon_{\mathcal{T}}(t) = \frac{\sum_{m=1}^k \sum_{i \in C_m} \sum_{j \notin C_m} S_{\mathcal{T}}^{i,j}}{\sum_{i \in \mathcal{U}} \sum_{j \in \mathcal{U}} S_{\mathcal{T}}^{i,j}}.
\end{equation}
A rising $\epsilon_{\mathcal{T}}(t)$ indicates that cohesive task sub-groups are being cut apart.
The three indicators live on incommensurable scales, as $\epsilon_{\mathcal{L}}$ is a squared distance ratio while $\epsilon_{\mathcal{T}}$ is a normalized cut ratio. Each is standardized by its own exponential moving scale $s_k(t) = \rho\, s_k(t-1) + (1-\rho)\, |\epsilon_k(t)|$, yielding the dimensionless defect $\hat{\epsilon}_k = \epsilon_k / s_k$. With $\hat{\boldsymbol{\epsilon}}(t) = [\hat{\epsilon}_{\mathcal{L}}, \hat{\epsilon}_{\mathcal{I}}, \hat{\epsilon}_{\mathcal{T}}]^T$, the weighting is posed as the online minimization of the aggregate normalized defect:
\begin{equation}
\min_{\boldsymbol{\beta} \in \Delta}\ \ J(\boldsymbol{\beta}) = \mathbf{1}^T \hat{\boldsymbol{\epsilon}}(\boldsymbol{\beta}),
\label{eq:beta_cost}
\end{equation}
\noindent where $\Delta = \{\boldsymbol{\beta} : \sum_{i=1}^3 \beta_i = 1,\ \beta_i \geq \beta_{\min}\}$ and the floor $\beta_{\min}$ keeps every modality alive. Minimizing $J$ steers the weight toward the modality that currently explains the swarm structure, since a modality that fails to do so inflates its own defect term.
As each $\epsilon_k$ depends on $\boldsymbol{\beta}$ only through the discrete clustering output, $J$ admits no closed-form gradient. We adopt projected gradient descent with a simultaneous perturbation stochastic approximation (SPSA) estimate \cite{spall1992multivariate}, which probes the cost along one random direction and needs two extra evaluations of $J$ per update:
\begin{equation}
\begin{aligned}
\hat{\mathbf{g}}(t) &= \frac{J\big(\Pi_{\Delta}(\boldsymbol{\beta} + c\,\mathbf{d})\big) - J\big(\Pi_{\Delta}(\boldsymbol{\beta} - c\,\mathbf{d})\big)}{2c}\,\mathbf{d},\\
\boldsymbol{\beta}(t) &= \Pi_{\Delta}\Big( \boldsymbol{\beta}(t-1) - \eta\, \hat{\mathbf{g}}(t) \Big),
\label{eq:beta_update}
\end{aligned}
\end{equation}
\noindent where $c$ is the perturbation magnitude, $\mathbf{d} \in \{-1,+1\}^3$ is a Rademacher vector, $\eta$ is the step size, and $\Pi_{\Delta}$ denotes Euclidean projection onto the simplex. The perturbation is applied once every $T_{\beta}$ slots to limit overhead. The recursion converges to a stationary point of $J$, at which no modality can be down-weighted without inflating the aggregate defect.
Coupling this feedback with the proposed FSC and the on-demand triggers completes the formation diverge and merge algorithm, given as Alg.~\ref{Alg:DynamicControl}.

\section{Cluster-Aware Distributed TDMA}
\label{sec:mac_protocol}

The diverge-merge algorithm in Section~\ref{sec:methodology} outputs spatial- and task-aware cluster labels. 
These labels indicate which UAVs should frequently exchange control information and task-related data. However, in dense corridors, formation diverge-merge maneuvers may suddenly increase network traffic load. If the MAC layer cannot support these communication bursts, UAVs may suffer from large access delay and frequent air-interface packet losses, finally affecting flight safety during formation maneuvers. Therefore, CAD-TDMA is designed as a network mechanism for formation diverge-merge control.

Regarding this, the MAC layer should support a large cluster size while keeping communication quality acceptable. Let $N_{\max}$ denote the maximum cluster size supported by the MAC layer. The objective is:
\begin{equation}
\begin{gathered}
\textbf{P2}: \ \ \max  \ \ N_{\max} \\
\mathrm{s.t.} \quad \bar{\Delta} \leq \Delta_{\mathrm{th}},  \!\!\!\!\!\!\!\!
\quad \quad \ \ell_{\mathrm{air}} \leq \ell_{\mathrm{th}}.
\end{gathered}
\label{eq:capacity_problem}
\end{equation}
where $\bar{\Delta}$ is the average packet delay and $\ell_{\mathrm{air}}$ is the air-interface loss rate. The air-interface loss rate represents failed MAC-layer delivery caused by simultaneous transmissions, slot conflicts, strong interference, or SINR degradation, and thus reflects the collision and interference risk at the air interface. This objective is feasibility-oriented rather than throughput-maximization-oriented: CAD-TDMA first aims to support the largest feasible cluster size under bounded delay and air-interface loss, while throughput $\Theta$ is used afterwards to evaluate slot-resource utilization. Therefore, the ns-3 evaluation in Section~\ref{sec:simulation} does not simply compare the maximum achievable throughput, but examines the delay-loss-throughput tradeoff under the same dynamic diverge-merge traces.

\begin{figure}[H]
\centering
\includegraphics[width=0.8\linewidth]{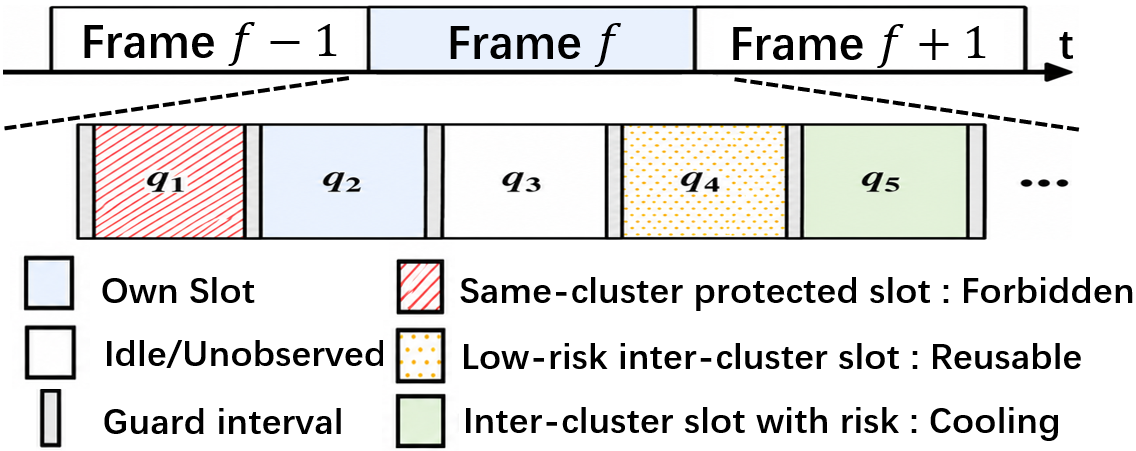}
\caption{Frame structure and representative slot states in CAD-TDMA.}
\label{fig:cad_frame_slot}
\end{figure}
The core idea of CAD-TDMA is to distinguish protected intra-cluster access from cautious inter-cluster reuse. Intra-cluster communication is usually more frequent and more important for diverge-merge synchronization, so same-cluster slots should be protected and not reused. Inter-cluster communication is relatively less coupled, so inter-cluster slots can be reused only when the distance and recent overhearing history indicate low risk. In this way, CAD-TDMA reduces delay and air-interface losses for critical intra-cluster traffic while maintaining effective resource utilization through conservative inter-cluster slot reuse.

\subsection{Frame-slot structure and state}

CAD-TDMA divides time into consecutive frames, and each frame contains $N_{\mathrm{slot}}$ configurable slots indexed by $\mathcal{Q}={1,2,\ldots,N_{\mathrm{slot}}}$. Each UAV maintains one owner slot $\omega_i(t)\in\mathcal{Q}$ as its primary reserved transmission opportunity in frame $t$. The owner slot is applied for reservation announcements, cluster-state control messages, and regular data packets. When the local queue becomes busy, a UAV may occupy low-risk inter-cluster slots for opportunity access.
For a single UAV, each slot is classified into one of five states, as shown in Fig.~\ref{fig:cad_frame_slot}.

\begin{itemize}[leftmargin=*]
\item \textbf{Own}: A slot owned by the node. The node can directly transmit data in this slots.
\item \textbf{Reusable}: A low-risk inter-cluster slot. The node can occupy it as an opportunity slot when the queue is busy.
\item \textbf{Forbidden}: An owner slot declared by another node. The node cannot occupy it.
\item \textbf{Cooling}: An inter-cluster slot with recent risk or insufficient spatial separation. The node should temporarily avoids it.
\item \textbf{Idle/Unobserved}: A slot without a fresh slot-binding record. It is designed as a candidate for future owner-slot allocation.
\end{itemize}

Therefore, a UAV can transmit only in its Own slot or in a Reusable slot. In Forbidden, Cooling, and Idle/Unobserved slots, the UAV keeps silent and listens to nearby packets. The guard interval between adjacent slots is only used to absorb timing offsets and is not treated as a transmission resource.

\subsection{Passive slot-view update}

\begin{figure}
\centering
\includegraphics[width=0.9\columnwidth]{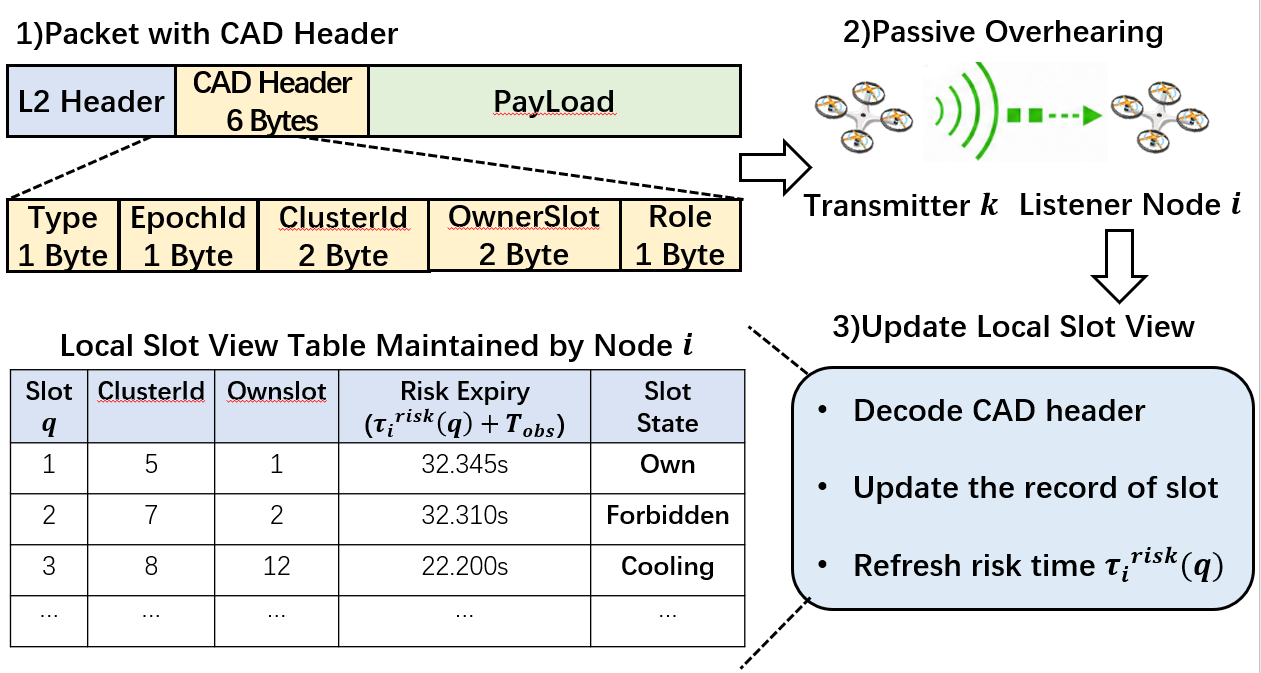}
\caption{Slot diagram of CAD-header-based passive update.}
\label{fig:cad_header}
\end{figure}

To support distributed operation, each CAD-TDMA packet carries a lightweight CAD header between the L2 header and the upper-layer payload, as illustrated in Fig.~\ref{fig:cad_header}. By passively overhearing ordinary packets, a UAV can decode the CAD header and update its local slot-view records, including the observed slot binding, cluster identity, freshness, and risk state. Therefore, CAD-TDMA does not require a centralized scheduler or an explicit reservation handshake in every frame.

The decoded CAD header provides the minimum information needed for local slot-view maintenance. Specifically, it contains five fields:
\begin{itemize}[leftmargin=*]
\item \textbf{Type}: packet type, such as data transmission, reservation announcement, release notification, or conflict feedback.
\item \textbf{EpochId}: cluster-state version. It helps discard stale slot-binding records after diverge or merge events.
\item \textbf{ClusterId}: sender's current cluster label. It is used to distinguish same-cluster protected slots from inter-cluster reuse candidates.
\item \textbf{OwnerSlot}: sender's reserved owner slot.
\item \textbf{Role}: sender's role, such as cluster head, boundary node, or ordinary member.
\end{itemize}

By overhearing CAD headers, a UAV can maintain a local slot view of nearby transmissions. In particular, it can determine which slots are occupied by same-cluster nodes, which slots belong to inter-cluster nodes, and whether the corresponding records are still fresh. In addition, undecodable busy slots, failed acknowledgements, and conflict feedback are recorded as risk evidence. These local observations are then used for next-frame slot classification and allocation.

\subsection{Cluster-aware allocation and inter-cluster reuse}

CAD-TDMA first protects same-cluster slots. Let $g_i(t)$ denote the cluster label of node $i$ in frame $t$, and let $\omega_i(t)$ denote its owner slot. After overhearing CAD headers, node $i$ builds a same-cluster blocked set:
\begin{equation}
\mathcal{B}_i(t)=\{\omega_j(t)\mid g_j(t)=g_i(t),~j\neq i,~\mathrm{Fresh}_j(t)=1\},
\label{eq:blocked_set}
\end{equation}
where $\mathrm{Fresh}_j(t)=1$ means that the record of node $j$ has a consistent EpochId and has not expired. Node $i$ must select its owner slot outside $\mathcal{B}_i(t)$, so same-cluster UAVs do not reuse each other's reserved slots.

This rule prevents UAVs in the same cluster from reusing each other's reserved slots. Since same-cluster UAVs frequently exchange heartbeat packets, acknowledgements, and cluster-state updates, such protection reduces high-risk simultaneous transmissions and helps lower access delay and air-interface losses.

After same-cluster slots are excluded, CAD-TDMA checks whether an inter-cluster slot can be reused. Let $\mathcal{P}_i(q,t)$ denote the set of fresh passive overhearing records associated with inter-cluster transmitters on slot $q$ at node $i$. The distance safety of slot $q$ is evaluated by the minimum distance from node $i$ to the transmitters in $\mathcal{P}_i(q,t)$. If no fresh inter-cluster record exists on slot $q$, the slot is treated as Idle/Unobserved rather than being directly reused.

Let $\tau_i^{\mathrm{risk}}(q)$ denote the latest frame in which node $i$ observes risk evidence on slot $q$, and let $T_{\mathrm{obs}}$ be the observation window. The available opportunity-slot set of node $i$ is defined as
\begin{equation}
\begin{aligned}
\mathcal{A}_i(t)=
\bigl\{q\in\mathcal{Q}\mid
&~q\notin\mathcal{B}_i(t),~
\mathcal{P}_i(q,t)\neq\emptyset,\\
&\mkern-145mu
\min_{j\in\mathcal{P}_i(q,t)}
\|p_i(t)-p_j(t)\|\ge \delta_{\mathrm{reuse}},~
t-\tau_i^{\mathrm{risk}}(q)>T_{\mathrm{obs}}
\bigr\},
\end{aligned}
\label{eq:reuse_rule}
\end{equation}
where the first condition protects same-cluster slots, the second ensures that the slot is a known inter-cluster slot rather than an unobserved slot, the third avoids reuse when the inter-cluster transmitter is too close, and the fourth requires that no recent risk evidence has been observed. A node may select opportunity slots only from $\mathcal{A}_i(t)$.

\begin{algorithm}
\caption{CAD-TDMA Slot-state Decision}
\label{Alg:CADTDMA}
\textbf{Input}: Cluster label $g_i(t)$, epoch, role, slot set $\mathcal{Q}$
\textbf{Output}: Owner slot $\omega_i(t)$ and opportunity-slot $\mathcal{A}_i(t)$
\For{each frame $f$}
{
Update local slot view by overhearing CAD headers

Remove stale records and update $\tau_i^{\mathrm{risk}}(q)$

Build $\mathcal{B}_i(t)$ according to (\ref{eq:blocked_set})

\If{$\omega_i(t)$ is invalid or $\omega_i(t)\in\mathcal{B}_i(t)$}
{
Select $\omega_i(t)\in\mathcal{Q}\setminus\mathcal{B}_i(t)$
}

Build $\mathcal{A}_i(t)$ according to (\ref{eq:reuse_rule})

\For{each slot $q$ in the next frame}
{
Transmit if $q=\omega_i(t)$ or $q\in\mathcal{A}_i(t)$ with a busy queue; otherwise overhear
}

\If{node $i$ exits or changes epoch}
{
Release or age the previous owner slot
}
}
\end{algorithm}

\subsection{Distributed protocol procedure}

\begin{figure}
\centering
\includegraphics[width=0.65\columnwidth]{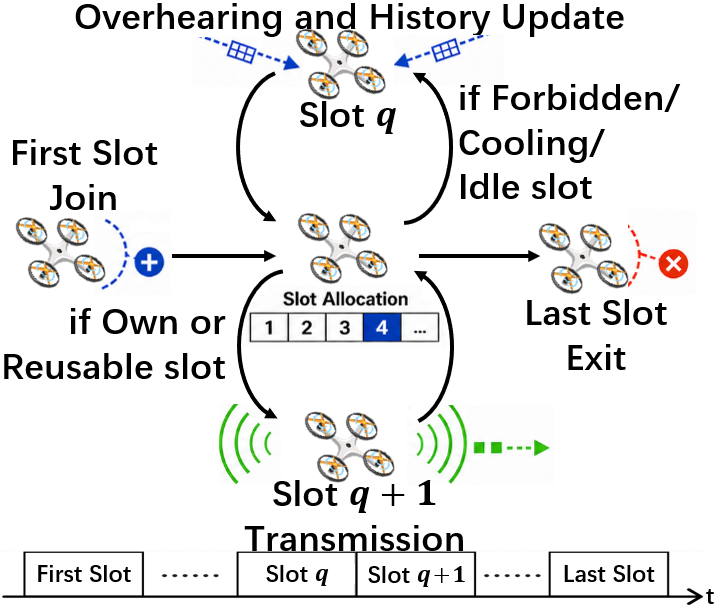}
\caption{Distributed CAD-TDMA protocol.}
\label{fig:cad_operation}
\end{figure}

The distributed protocol of CAD-TDMA is summarized in Fig.~\ref{fig:cad_operation} and Alg.~\ref{Alg:CADTDMA}. Each UAV maintains a local slot view by overhearing CAD headers in the slots where it does not transmit. At the beginning of each frame, stale records are removed according to the EpochId and aging timer. The node then updates its same-cluster blocked set and evaluates whether inter-cluster slots have sufficient spatial separation and no recent risk evidence.

If the current owner slot becomes invalid due to a cluster-epoch change or a same-cluster conflict, the UAV selects a new owner slot outside the same-cluster blocked set. For data transmission, the node first uses its owner slot. When the queue is busy, it may occupy slots classified as reusable. In Forbidden, Cooling, and Idle/Unobserved slots, the node remains silent and continues passive listening. When a node leaves or changes its cluster epoch, its previous slot binding is removed either by explicit release or by local aging.













\section{Numerical Simulation}
\label{sec:simulation}

This section presents a comprehensive numerical validation of the proposed dynamic diverge-merge collaborative control framework. The evaluation methodology is structured into two primary components: 1) a {kinematic algorithmic validation} using a Python-based simulator to assess the topological reconstruction accuracy of a 100-UAV swarm; and 2) a {communication-support validation} using ns-3 to evaluate the underlying MAC protocol's capability in supporting high-frequency topology evolution.

\subsection{Diverge-Merge Algorithm Validation}
\label{sec:dm_validation}

To evaluate the proposed framework, we implement a 3D kinematic UAV simulator. The mobility follows the APF-based bounded control model established in Section~\ref{sec:system_model}, while the communication layer evaluates dynamic SINR using a log-distance path-loss model. The primary simulation parameters are summarized in Table~\ref{tab:sim_params}. 

\begin{table}[htbp]
\centering
\caption{Primary simulation parameters}
\label{tab:sim_params}
\begin{tabular}{l l | l l}
\toprule
\textbf{Param.} & \textbf{Value} & \textbf{Param.} & \textbf{Value} \\
\midrule
$N$          & $100$        & $\Delta t$   & $0.1$~s \\
$T_{c}$      & $1$~s        & $v_{\max}$   & $15$~m/s \\
$a_{\max}$   & $3$~m/s$^2$  & $d_0$        & $5$~m \\
$\kappa_0$   & $60$~m/s$^2$ & $R_c$        & $250$~m \\
$P_{tx}$     & $23$~dBm     & $\alpha$     & $2.5$ \\
$T_w$        & $400$~s      & $N_{\max}$   & $45$ \\
$d_{merge}$  & $45$~m       & $d_{diverge}$ & $90$~m \\
\bottomrule
\end{tabular}
\end{table}

We benchmark the proposed framework against two baselines. The proposed framework executes the complete pipeline equipped with adaptive weights and TD-FPTree task memory. Standard SC represents spectral clustering driven only by spatial and intent similarities ($S_{\mathcal{L}}$ and $S_{\mathcal{I}}$), lacking historical task memory. K-Means serves as the spatial-only baseline, applying standard K-means directly to raw 3D coordinates.

To quantitatively capture the utility of the resulting topologies, we define two outcome-oriented metrics based on the ground-truth task groups $\{\mathcal{T}_g\}_{g=1}^G$ and the algorithm-assigned clusters $\{C_i\}_{i=1}^N$. 

First, the Task-Cluster Alignment (TCA) measures the proportion of same-task UAVs successfully grouped into the same macroscopic cluster. It is defined as the ratio of the most frequent cluster label within each task group:
\begin{equation}
    \text{TCA} = \frac{1}{G} \sum_{g=1}^G \frac{1}{|\mathcal{T}_g|} \max_c \sum_{i \in \mathcal{T}_g} \mathbb{I}(C_i = c),
\end{equation}
where $\mathbb{I}(\cdot)$ is the indicator function.

Second, the Task Communication Score (TCS) evaluates the effective data exchange quality among collaborating nodes. Intra-cluster links preserve their full capacity, whereas cross-cluster links suffer a penalty to model multi-hop routing overhead:
\begin{equation}
    \text{TCS} = \!\!\frac{1}{N_{pair}} \!\!\sum_{g=1}^G \sum_{i < j \in \mathcal{T}_g} \!\!\!S_{\mathcal{L}}^{i,j} \Big( \mathbb{I}(C_i = C_j) + \delta \mathbb{I}(C_i \neq C_j) \Big),
\end{equation}
where $N_{pair}$ is the total number of intra-task UAV pairs, and $\delta = 0.02$ represents the severe multi-hop penalty.

Two stress-test scenarios are designed to expose the limitations of conventional algorithms. The \textit{Multi-Corridor scenario} places five heterogeneous task groups in a network of four parallel corridors in the lower layer and one corridor in the upper layer, each group starting in its own corridor. During $t\in[25,50]$~s, two groups enter the main lower corridor through ramps from opposite sides while a third climbs a ramp into the upper corridor, so that same-layer and cross-layer merging occur together and create severe physical overlapping, before differential-speed peeling restores the formations.

The \textit{Congestion scenario} confines five fleets to a single multi-lane corridor, where they enter in an ordered queue and are initially well separated. A downstream bottleneck decelerates the leading fleet during $t\in[20,50]$~s, and the resulting shockwave propagates backward, compressing the queue to near-collision spacing and driving the fleets to interpenetrate. Several fleets share the same downstream destination, so the declared intent distinguishes traffic only at the ramp level rather than at the fleet level. Spatial distance and flight intent therefore lose discriminative power as a consequence of congestion physics rather than by construction.

\begin{figure}[t]
\centering
\includegraphics[width=\columnwidth]{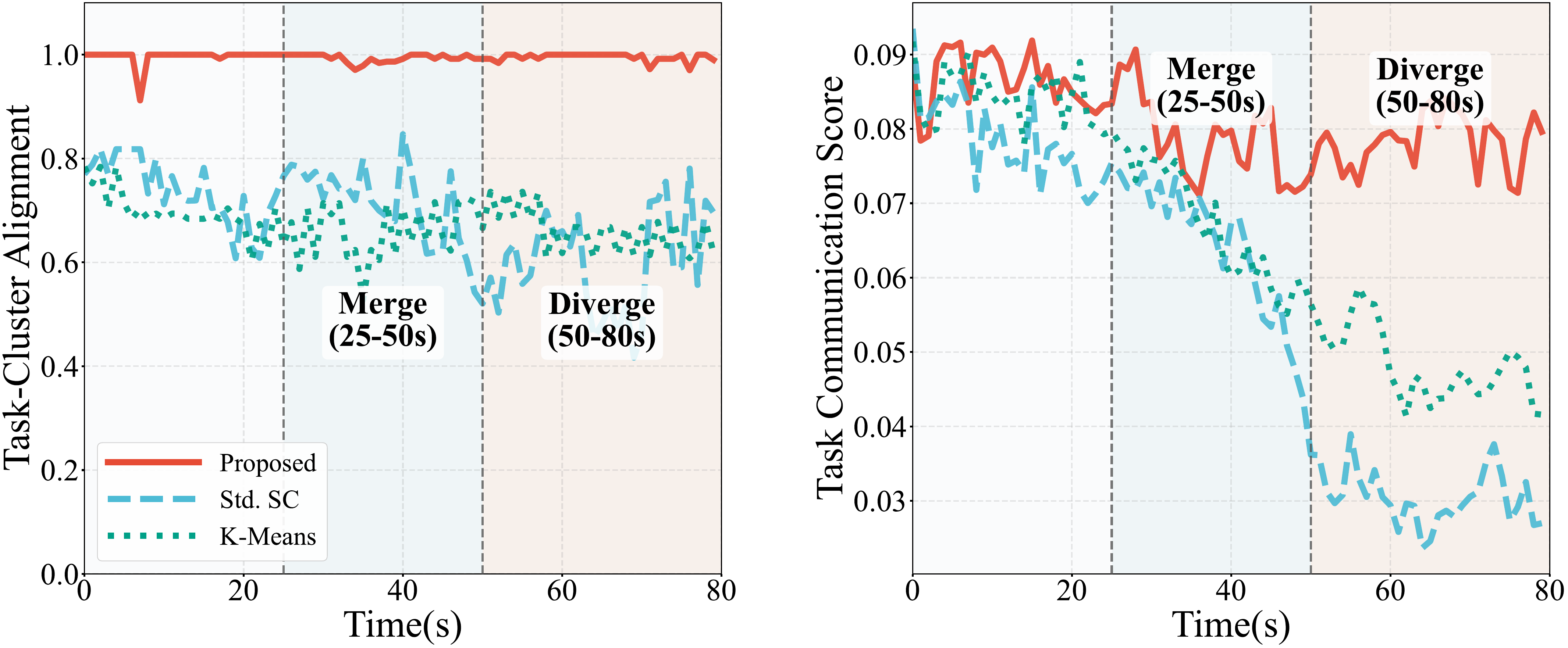}
\caption{Multi-Corridor scenario. The proposed framework sustains a TCA above $0.99$ across all phases, whereas both baselines degrade once the ramp maneuvers collapse the spatial separation.}
\label{fig:s1_metrics}
\end{figure}

\begin{figure}[t]
\centering
\includegraphics[width=\columnwidth]{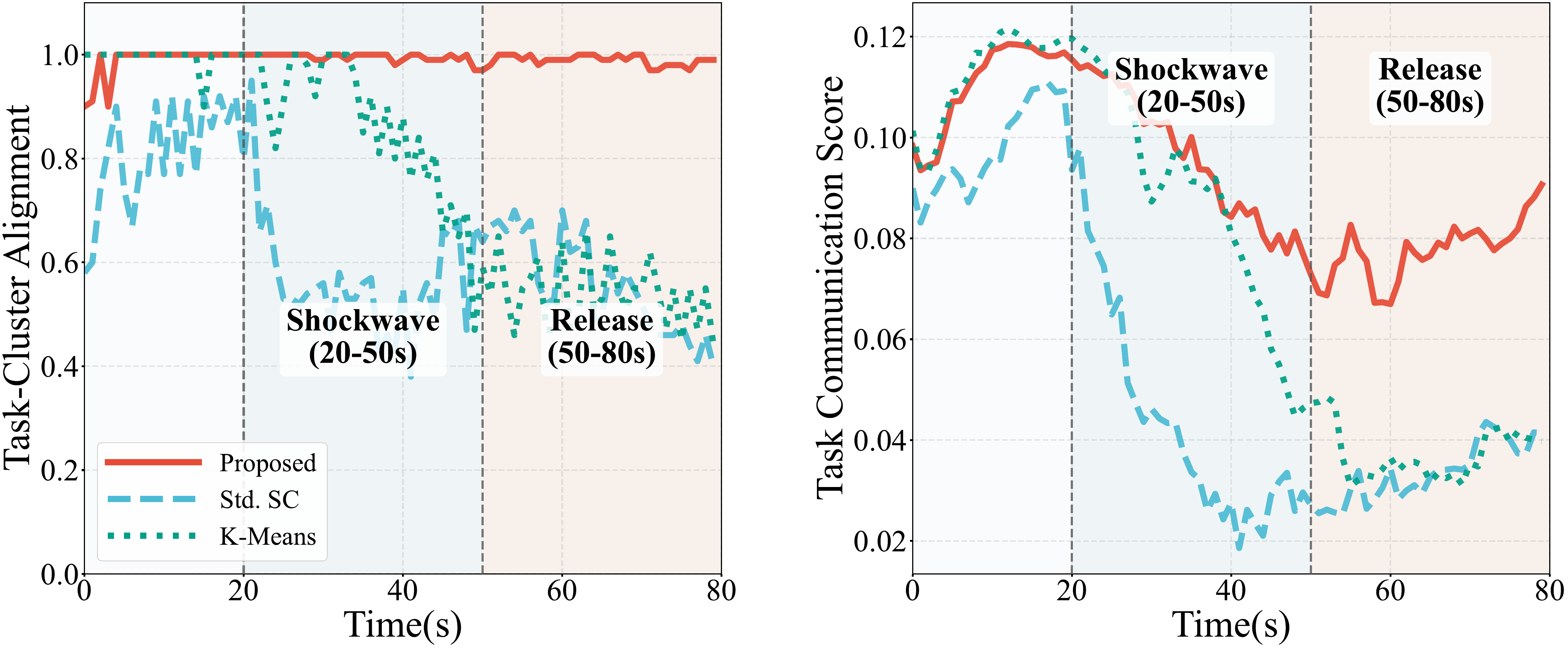}
\caption{Congestion scenario. Both baselines track the proposed framework under free flow, but fall away once the shockwave drives the fleets to interpenetrate.}
\label{fig:s2_metrics}
\end{figure}

\begin{table}[htbp]
\centering
\caption{TCA and TCS performance comparison}
\label{tab:mean_metrics}
\begin{tabular}{l | c c | c c}
\toprule
\textbf{Algorithm} & \multicolumn{2}{c|}{\textbf{Multi-Corridor (S1)}} & \multicolumn{2}{c}{\textbf{Congestion (S2)}} \\
& \textbf{TCA} & \textbf{TCS} & \textbf{TCA} & \textbf{TCS} \\
\midrule
Proposed & $\mathbf{0.995}$ & $\mathbf{0.081}$ & $\mathbf{0.991}$ & $\mathbf{0.093}$ \\
Std. SC  & $0.679$ & $0.056$ & $0.625$ & $0.054$ \\
K-Means  & $0.672$ & $0.066$ & $0.777$ & $0.075$ \\
\bottomrule
\end{tabular}
\end{table}

\begin{table}[htbp]
\centering
\caption{Mean cluster number $k$ per phase}
\label{tab:k_evolution}
\begin{tabular}{l | c c c | c c c}
\toprule
\textbf{Algorithm} & \multicolumn{3}{c|}{\textbf{Multi-Corridor (S1)}} & \multicolumn{3}{c}{\textbf{Congestion (S2)}} \\
& Indep. & Merge & Diverge & Free & Shock & Release \\
\midrule
Proposed & $\mathbf{5.05}$ & $\mathbf{5.00}$ & $\mathbf{5.00}$ & $\mathbf{5.00}$ & $\mathbf{5.00}$ & $\mathbf{5.00}$ \\
Std. SC  & $4.35$ & $4.20$ & $3.73$ & $5.87$ & $4.07$ & $4.23$ \\
K-Means  & $7.50$ & $5.92$ & $4.97$ & $5.07$ & $4.67$ & $4.00$ \\
\bottomrule
\end{tabular}
\end{table}

As summarized in Table~\ref{tab:mean_metrics}, the proposed framework attains the highest TCA and TCS in both scenarios. Table~\ref{tab:k_evolution} reports the cluster number resolved by each algorithm, which is not supplied as a prior but is decided online by the eigengap estimate under the capacity bound and the diverge-merge triggers. This quantity exposes a failure mode that TCA alone cannot reveal, since a partition that absorbs several task groups into one cluster still retains a high alignment score.

In the Multi-Corridor scenario, the proposed framework recovers $k \approx 5$ in every phase and sustains a TCA above $0.99$ throughout. Fig.~\ref{fig:s1_metrics} shows the transient evolution. Even before any maneuver, K-Means resolves $k = 7.50$, because a formation extended along its corridor is split into leading and trailing groups by a purely spatial partition, and its TCA is already limited to $0.682$. During the merging phase ($t \in [25,50]$~s), the two same-layer groups entering the main corridor and the group climbing into the upper layer overlap heavily in space, and the collapsed spatial separation leaves the spectral embedding of Std. SC without a discriminative cut, so its TCA falls to $0.716$. In the diverge phase, differential-speed peeling restores the spatial separation, yet neither baseline recovers: Std. SC drifts to $k = 3.73$ and a TCA of $0.581$, absorbing distinct groups into common clusters, while its TCS declines to $0.030$. The proposed framework retains the group structure through the task-affinity modality, whose historical interaction record is unaffected by the transient spatial overlap.

In the Congestion scenario, all fleets share one corridor and the shockwave compresses the queue to near-collision spacing. Fig.~\ref{fig:s2_metrics} traces the transient evolution. Under free flow the fleets remain well separated, and K-Means attains a TCA of $0.955$ with a near-correct $k=5.07$, which confirms that the scenario does not penalize spatial reasoning by construction. Once the shockwave arrives, the fleets interpenetrate and spatial proximity ceases to be informative: K-Means falls to $0.787$ and Std. SC to $0.571$, whereas the proposed framework holds at $0.993$. After the bottleneck is released, both baselines collapse further rather than recovering. K-Means settles at $k = 4.00$ with a TCA of $0.452$, and Std. SC at $k = 4.23$ with a TCA of $0.558$, since several fleets exit through the same downstream ramp and the declared intent no longer separates them. The proposed framework maintains $k = 5.00$ and a TCA of $0.986$, and reorganizes the fleets onto their own ramps.

\begin{figure}[t]
\centering
\includegraphics[width=0.7\columnwidth]{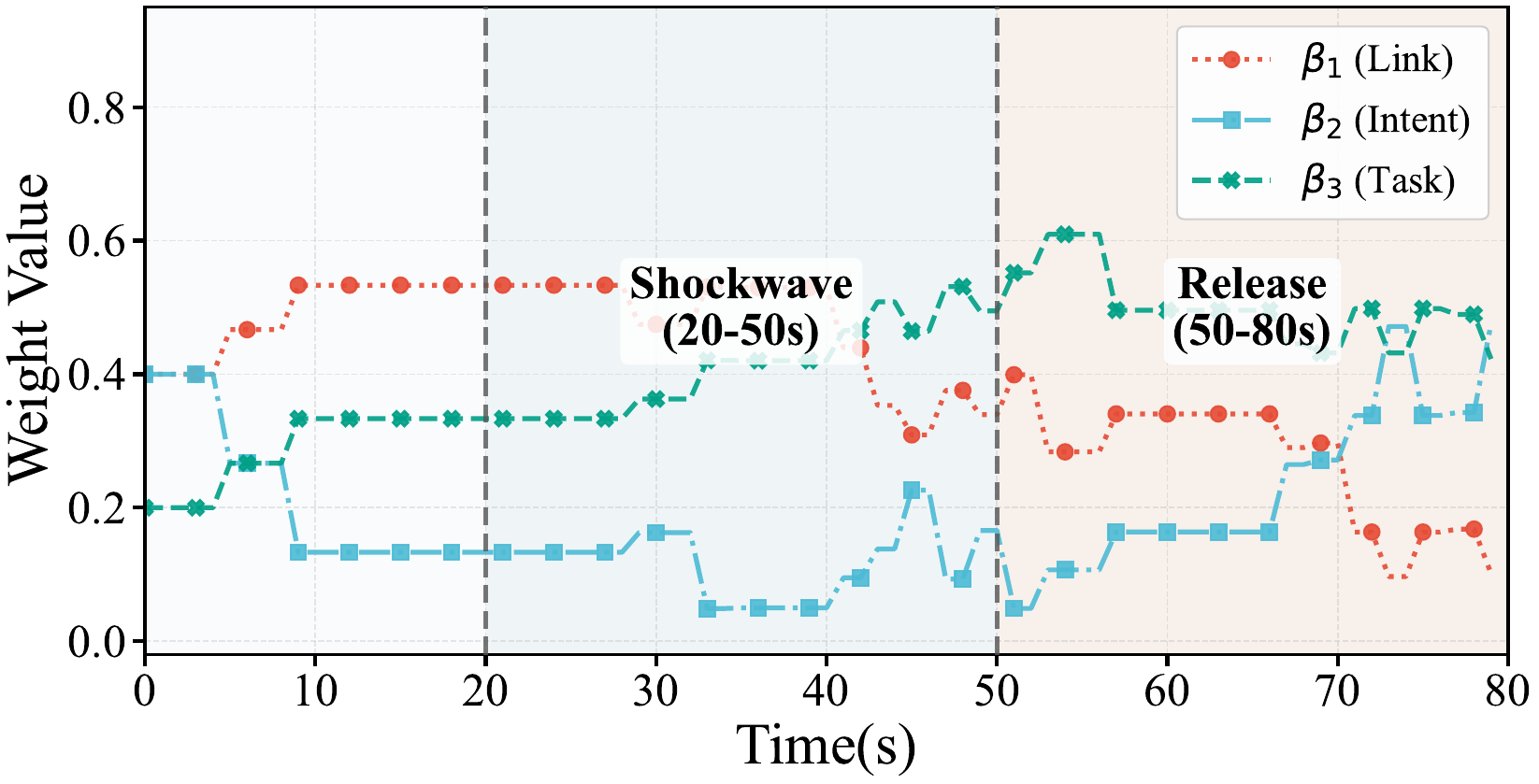}
\caption{Dynamic weight adaptation in the congestion scenario. The task weight $\beta_3$ rises as the shockwave removes the spatial signal.}
\label{fig:beta_adaptation}
\end{figure}

Fig.~\ref{fig:beta_adaptation} traces the weight adaptation in the Congestion Scenario and reveals the underlying closed-loop mechanism. Under free flow the link modality takes the lead as the well-separated fleets make spatial proximity the most reliable cue, while the intent weight is suppressed early. As the shockwave removes the spatial signal, the weight shifts toward the task modality, which rises to share the lead with the link term and carries the fleet partition that neither position nor destination can recover. This task-dominant regime persists after the release, confirming that the controller tracks the currently informative modality rather than a fixed weighting.

\begin{figure}[t]
\centering
\includegraphics[width=.7\columnwidth]{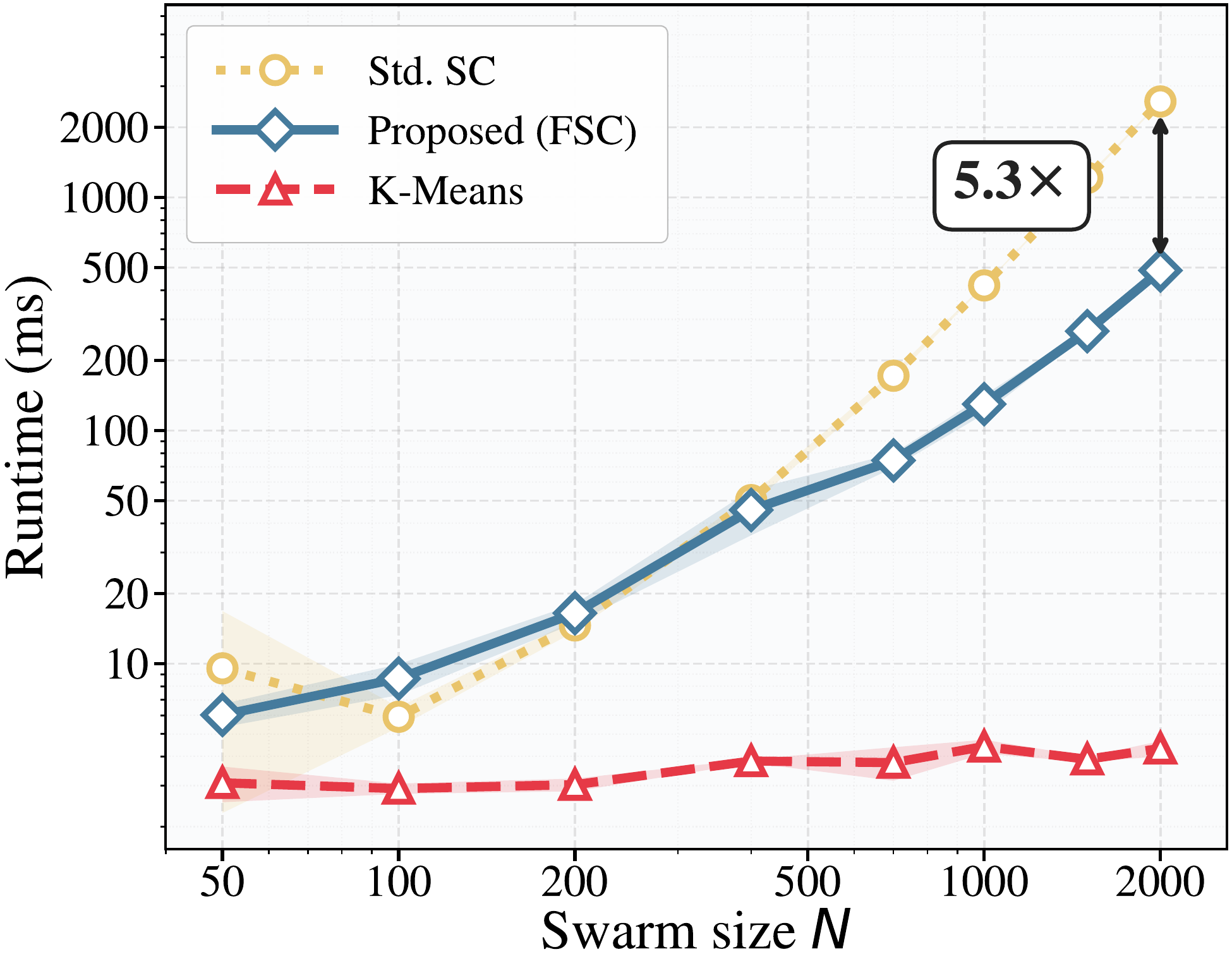}
\caption{Execution time comparison across varying formation sizes.}
\label{fig:complexity}
\end{figure}

Finally, Fig.~\ref{fig:complexity} validates the computational scalability of the proposed framework. Across formation sizes $N \in [50, 2000]$, it completes the $N=2000$ partition in $0.49$~s against $2.62$~s for standard spectral clustering, a $5.3\times$ speedup that widens monotonically with $N$. The \textbf{total runtime} is dominated by the $O(N^2)$ similarity construction, as pairwise SINR must be evaluated for all links; however, the \textbf{FSC core} reduces the subsequent eigen decomposition from $O(N^3)$ to $O(Nk^2)$, which is the primary source of the observed performance gain. K-Means is faster in absolute terms but resolves an incorrect cluster number in every phase, as reported in Table~\ref{tab:k_evolution}.

\subsection{Communication-support validation with ns-3}

To verify whether the proposed diverge-merge results can be supported at the wireless access layer, we conduct a trace-driven ns-3 validation~\cite{riley2010ns3}. The ns-3 simulator replays the Python-side dynamic diverge-merge control framework traces, including UAV trajectories, task-aware cluster labels, MAC interaction events, and task-level traffic events. Two scenarios are evaluated: multi-corridor scenario for UAV corridor-switching and congestion scenario for extreme single-corridor congestion. 
Each scenario contains 80 replay frames with an interval of 1 s, resulting in a 79.5 s effective evaluation duration after warm-up and ending margins.

\begin{figure}[t]
\centering

\begin{subfigure}[t]{\linewidth}
\centering
\includegraphics[width=0.85\linewidth]{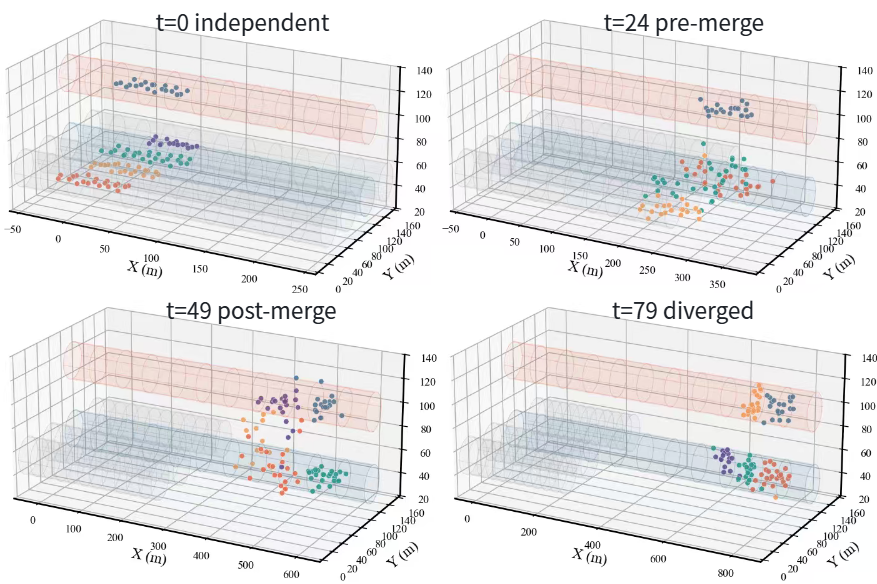}
\caption{Multi-corridor scenario.}
\label{fig:pos_s1}
\end{subfigure}

\vspace{1mm}

\begin{subfigure}[t]{\linewidth}
\centering
\includegraphics[width=0.85\linewidth]{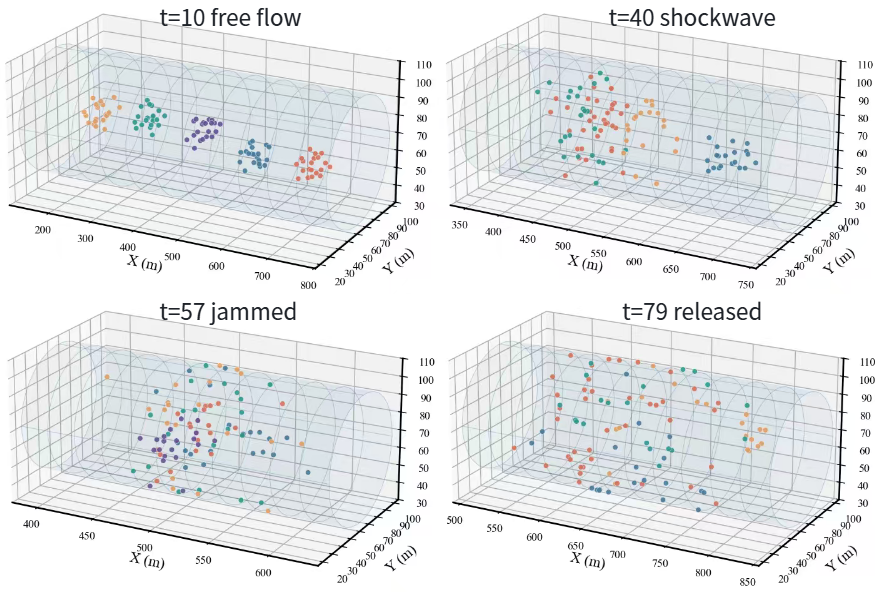}
\caption{Congestion scenario.}
\label{fig:pos_s2}
\end{subfigure}

\vspace{1mm}

\begin{subfigure}[t]{\linewidth}
\centering
\includegraphics[width=0.85\linewidth]{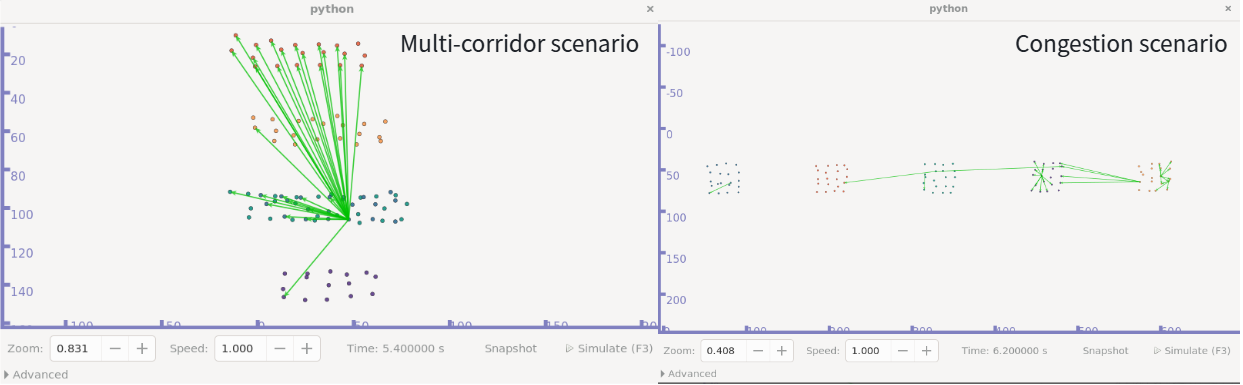}
\caption{NS-3 network connection results.}
\label{fig:ns3_replay}
\end{subfigure}

\caption{Scenario position snapshots and ns-3 replay results under two diverge-merge scenarios. Different colors indicate different task-aware clusters, and green arrows indicate the traffic flows.}
\label{fig:ns3_pyviz_snapshots}
\vspace{-1mm}
\end{figure}

Fig.~\ref{fig:ns3_pyviz_snapshots} shows the scenario position snapshots and ns-3 replay results for the multi-corridor scenario and the congestion scenario. The position snapshots illustrate the UAV spatial distributions during independent flight, merging, congestion, release, and diverge phases, while the ns-3 replay visualizes the corresponding network state, task-aware cluster labels, and traffic flows. CAD-TDMA uses these replayed labels for cluster-aware slot management, whereas WiFi and fixed TDMA are evaluated under the same traffic traces as baselines.

Control traffic is modeled by UDP heartbeat broadcasts with short ACK replies, while task traffic is modeled by burst UDP incast toward an event sink. The key simulation parameters are summarized in Table~\ref{tab:ns3_parameters}.

Three metrics are used: average delay $\bar{\Delta}$, air-interface loss rate $\ell_{\rm air}$, and throughput $\Theta$. Delay reflects the waiting time for cluster-state synchronization and bursty task traffic. Air-interface loss measures failed MAC-layer delivery after transmission and captures collision/interference risk. Throughput evaluates the effective use of wireless resources.

\begin{table}[t]
\centering
\caption{Primary ns-3 parameters.}
\label{tab:ns3_parameters}
\small
\setlength{\tabcolsep}{4pt}
\renewcommand{\arraystretch}{0.95}
\begin{tabular}{p{0.16\textwidth} p{0.28\textwidth}}
\hline
\textbf{Parameter} & \textbf{Value} \\
\hline
Replay setting & 80 frames, 1 s/frame, 79.5 s evaluation \\
Mobility input & Waypoint replay from dynamic diverge-merge collaborative control traces \\
Compared MACs & WiFi MAC, fixed TDMA, CAD-TDMA \\
PHY & IEEE 802.11a, 6 Mbps\\
Channel model & log-distance path loss \\
Traffic model & UDP heartbeat/ACK and burst incast \\
Packet size & 64-byte control, 800-byte data \\
TDMA setting & One-node-one-slot, 6 ms slot \\
\hline
\end{tabular}
\vspace{-1mm}
\end{table}

\begin{figure}[!t]
\centering

\begin{subfigure}[b]{0.49\linewidth}
\centering
\includegraphics[width=\linewidth]{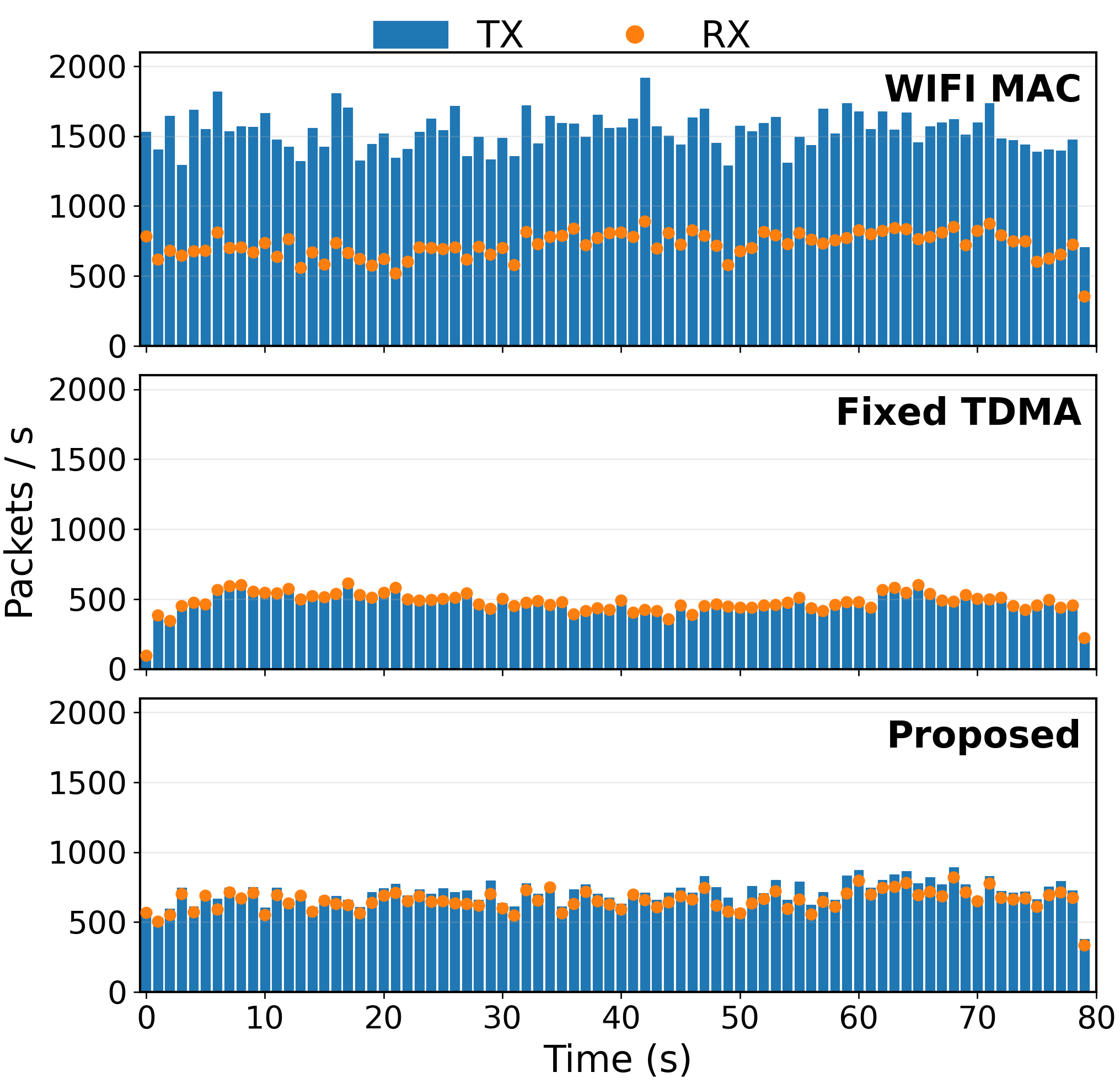}
\caption{Multi-corridor scenario.}
\label{fig:s1_txrx_dynamics}
\end{subfigure}
\hfill
\begin{subfigure}[b]{0.49\linewidth}
\centering
\includegraphics[width=\linewidth]{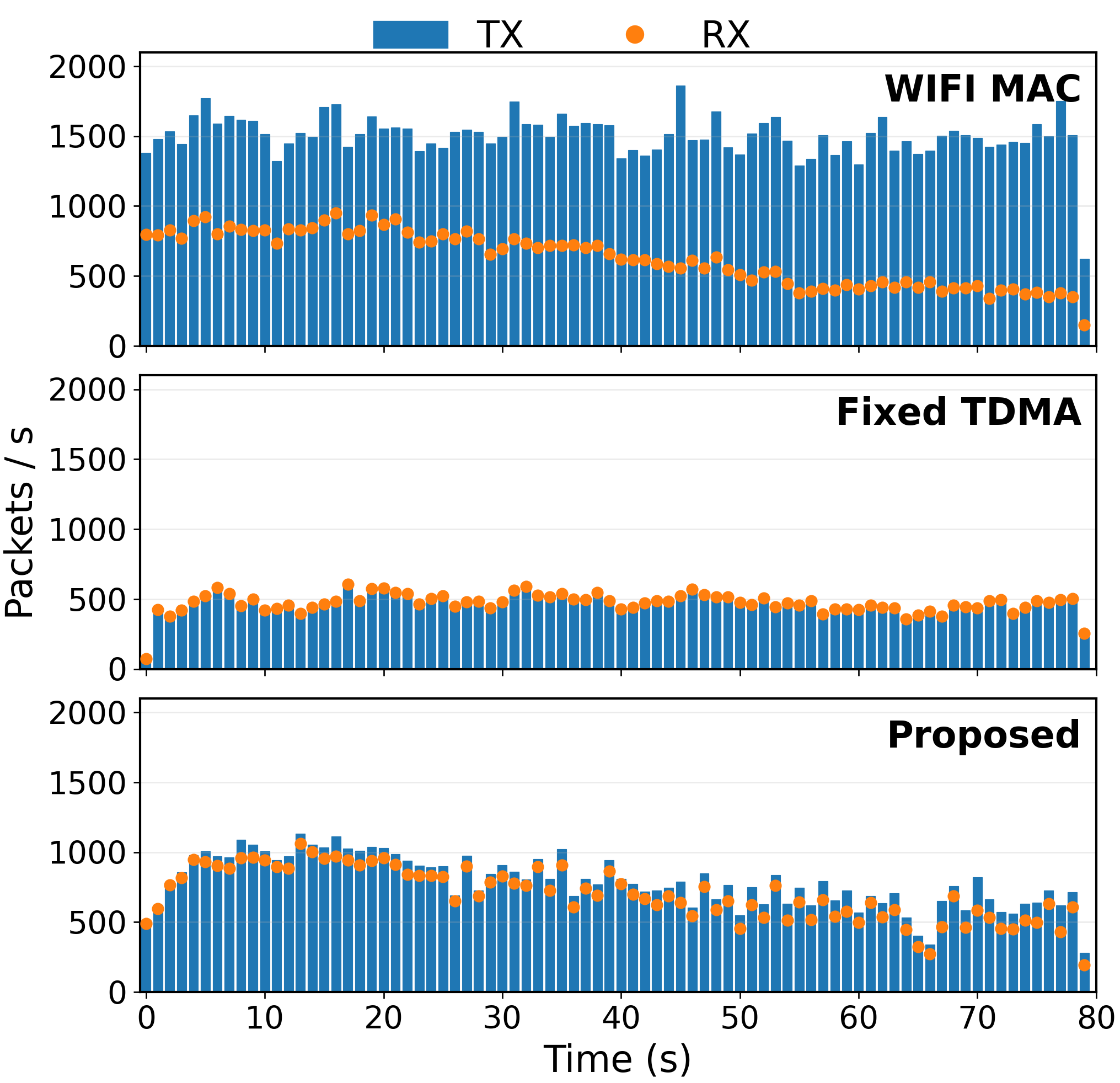}
\caption{Congestion scenario.}
\label{fig:s2_txrx_dynamics}
\end{subfigure}

\caption{Per-second TX/RX packet dynamics under two representative scenarios. In each subfigure, the three panels from top to bottom correspond to WiFi MAC, fixed TDMA, and Proposed.}
\label{fig:txrx_dynamics}
\vspace{-1mm}
\end{figure}

Specifically, $\bar{\Delta}= \frac{1}{N_{\rm rx}}\sum_{p\in\mathcal{P}_{\rm rx}}\left(t_p^{\rm rx}-t_p^{\rm tx}\right)$, where $\mathcal{P}_{\rm rx}$ is the set of received packets and $N_{\rm rx}=|\mathcal{P}_{\rm rx}|$. The air-interface loss rate is $\ell_{\rm air}= {N_{\rm air}^{\rm loss}}/{N_{\rm mac}^{\rm tx}}$, where $N_{\rm air}^{\rm loss}$ is the number of packets lost after MAC transmission and $N_{\rm mac}^{\rm tx}$ is the number of MAC-layer transmitted packets. The throughput is $\Theta= {8B_{\rm rx}}/{T_{\rm eval}}$, where $B_{\rm rx}$ is the total number of received bytes and $T_{\rm eval}$ is the evaluation duration. Therefore, the TX/RX dynamics in Fig.~\ref{fig:s1_txrx_dynamics} are used only to illustrate runtime access behavior, while the final comparison is based on delay, air-interface loss, and throughput jointly.






Fig.~\ref{fig:txrx_dynamics} shows the runtime TX/RX packet dynamics under the multi-corridor scenario and the congestion scenario with a unified vertical scale. WiFi produces the largest number of transmission attempts, but the clear TX--RX gap indicates considerable ineffective access under contention. Fixed TDMA keeps TX and RX close due to deterministic slot scheduling, while Proposed achieves a similarly balanced TX/RX behavior with more adaptive access control. This indicates that Proposed avoids excessive air-interface attempts and reduces communication-resource waste, while maintaining practical service capability under dynamic UAV formation traces.

\begin{figure}[!t]
\centering
\includegraphics[width=0.7\linewidth]{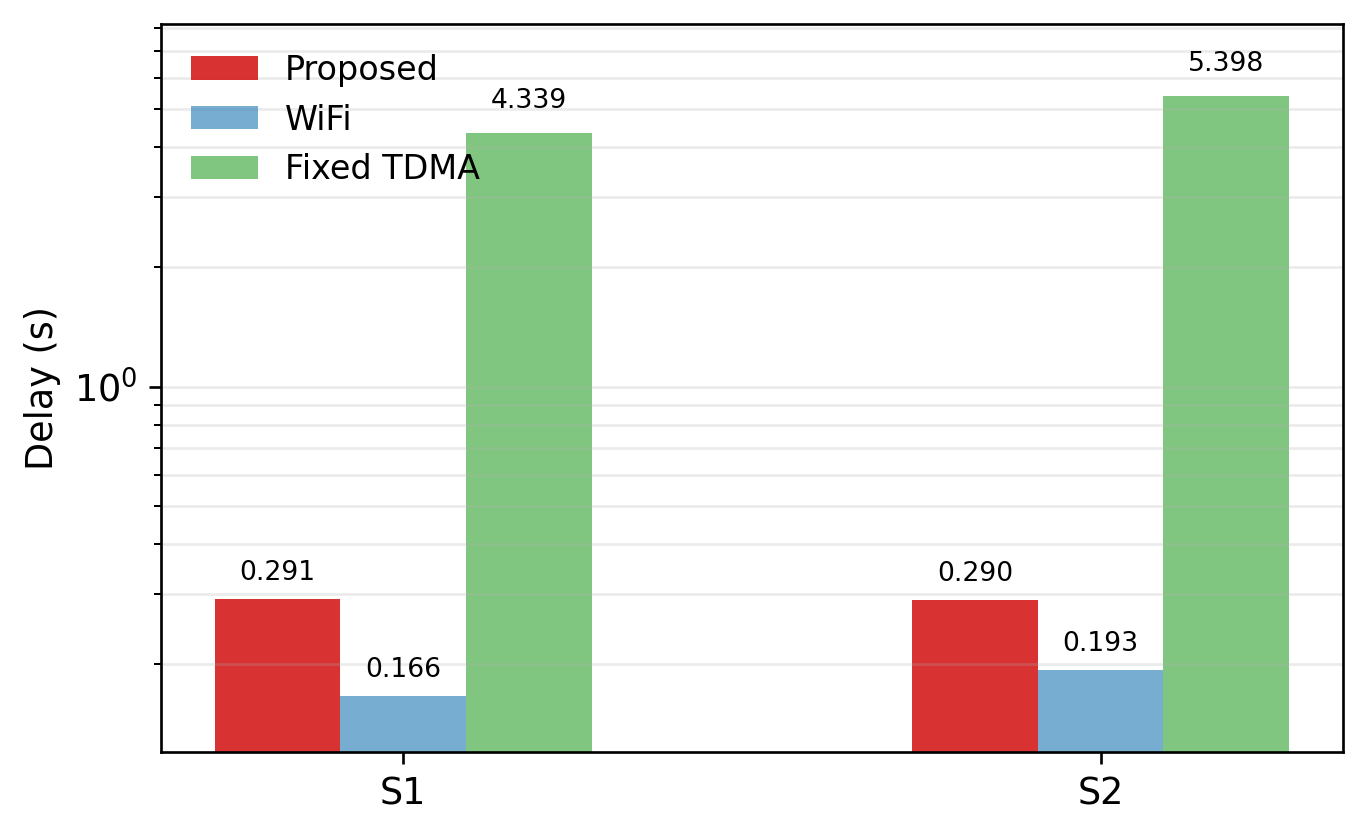}
\caption{Average delay comparison of different MAC schemes under multi-corridor scenario and congestion scenario.}
\label{fig:delay_bar}
\vspace{-1mm}
\end{figure}

Fig.~\ref{fig:delay_bar} compares the average delay in the two scenarios. WiFi has the smallest delay among successfully received packets because random access allows immediate transmission attempts. However, this apparent delay advantage is achieved at the cost of severe packet loss. Fixed TDMA avoids most air-interface collisions, but its rigid slot schedule introduces second-level waiting delay, which is unsuitable for frequent cluster-state updates and bursty task traffic. CAD-TDMA significantly reduces the delay compared with fixed TDMA by allowing cluster-aware slot management and low-risk opportunity-slot reuse.

\begin{table}[!t]
\centering
\caption{Level mapping for communication-support capability.}
\label{tab:capability_levels}
\scriptsize
\setlength{\tabcolsep}{3.2pt}
\renewcommand{\arraystretch}{1.08}
\begin{tabular}{c c c c}
\hline
\textbf{Level} & \textbf{Delay support(ms)} & \textbf{Reliability} & \textbf{Throughput (Mbps)} \\
\hline
5 & $\leq 200$ & $\leq 0.01$ & $\geq 6$ \\
4 & $(200,350]$ & $(0.01,0.10]$ & $[4,6)$ \\
3 & $(350,1000]$ & $(0.10,0.30]$ & $[2,4)$ \\
2 & $(1000,3000]$ & $(0.30,0.50]$ & $[1,2)$ \\
1 & $>3000$ & $>0.50$ & $<1$ \\
\hline
\end{tabular}
\vspace{-1mm}
\end{table}

\begin{figure}[!t]
\centering
\includegraphics[width=0.65\linewidth]{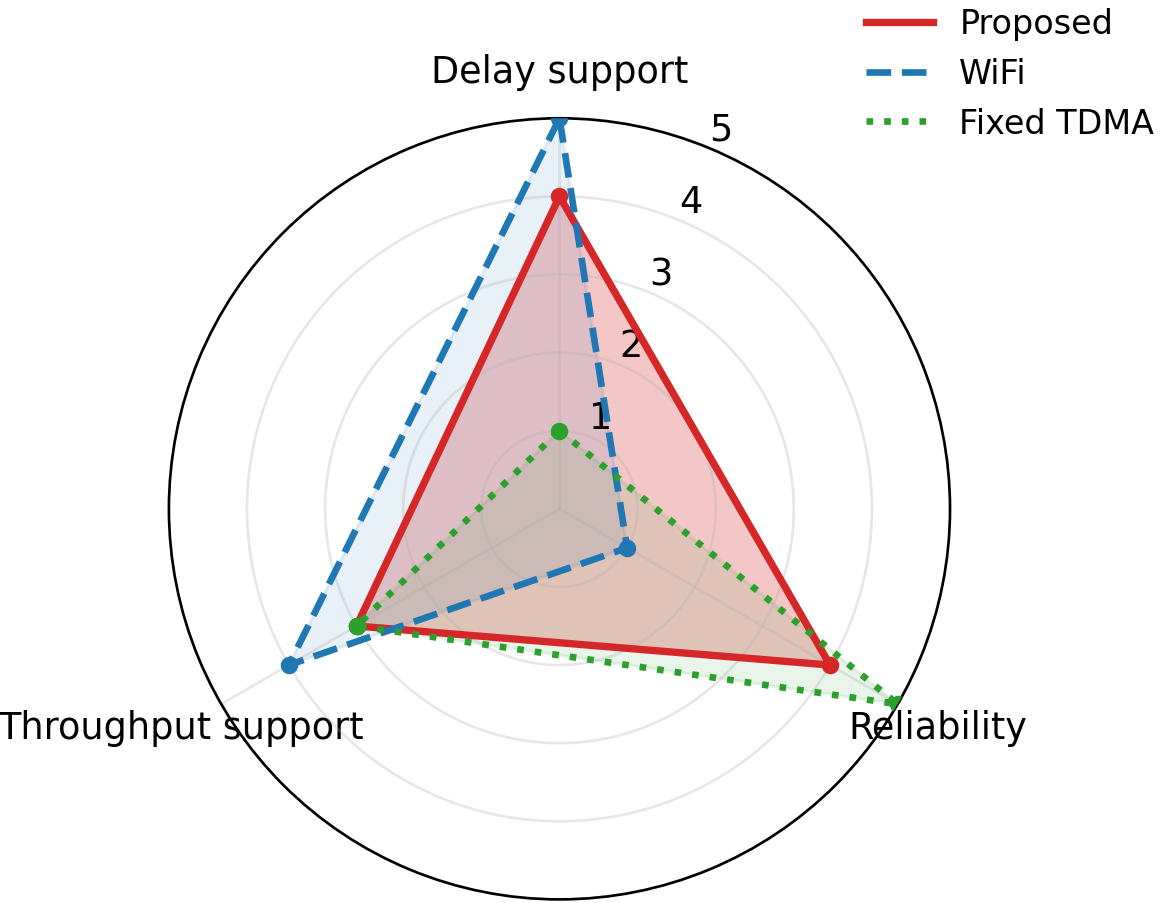}
\caption{Capability of WiFi MAC, fixed TDMA, and proposed.}
\label{fig:capability_radar}
\vspace{-1mm}
\end{figure}
For intuitive comparison, Table.~\ref{tab:capability_levels} indicates that the three communication metrics are mapped into fixed threshold ranges, where a higher level indicates better communication support. This level mapping is used only for visualization and does not introduce an additional optimization objective.

Fig.~\ref{fig:capability_radar} summarizes the communication-support capability in three aspects: delay support, reliability, and throughput support. A higher level indicates a better support capability. WiFi has strong delay and throughput support, but weak reliability because dense contention causes many air-interface losses. Fixed TDMA has strong reliability but weak delay support because its rigid schedule limits responsiveness. CAD-TDMA does not maximize a single metric; instead, it provides the most balanced capability profile. As observed in Fig.~\ref{fig:s1_txrx_dynamics}, CAD-TDMA may suppress transmissions during dense merge-overlap intervals, but this behavior reduces ineffective access and protects intra-cluster synchronization. Therefore, its throughput is acceptable rather than maximal, while its delay and reliability remain suitable for dynamic formation control.

 \section{Conclusion}

This paper addresses the dynamic diverge-merge control of UAV flight formations within corridor-ramp structured airspace, seamlessly bridging formation reconstruction with wireless access scheduling. To overcome the severe spatial squeezing and communication bottlenecks inherent in complex ramp transitions, we developed a task-driven collaborative control algorithm that fuses spatial connectivity, 3D flight intent, and historical task interactions via an accelerated spectral clustering paradigm, effectively preventing blind fragmentation under extreme physical congestion. 
Furthermore, to sustain the bursty network traffic demands of high-frequency formation evolution, a cluster-aware distributed TDMA protocol was formulated to protect intra-cluster owner slots while conservatively exploiting low-risk inter-cluster opportunities. Comprehensive kinematic and ns-3 trace-driven validations confirm that the proposed framework maintains near-zero network misclassification and achieves a superior delay-loss-throughput tradeoff over conventional baselines. Ultimately, this work provides a scalable, network-enabled blueprint for safe and high-capacity AAM operations.

\appendices
\footnotesize

\bibliography{biblio}
\end{document}